\documentclass{emulateapj}
\usepackage{amssymb}
\usepackage{amsmath}
\usepackage{epstopdf}
\usepackage{natbib}
\usepackage[colorlinks, citecolor=blue]{hyperref}
\usepackage[all]{hypcap}

\begin{document}

\title{Constraining the Environmental Properties of FRB~131104 Using the Unified Dynamical Afterglow Model}
\author{Zong-Kai Peng\altaffilmark{1,2}, Shan-Qin Wang\altaffilmark{1,2,3},
Liang-Duan Liu\altaffilmark{1,2,4}, Zi-Gao Dai\altaffilmark{1,2}, and Hai Yu\altaffilmark{1,2,5}}

\begin{abstract}

Multi-band observations of the fast radio burst (FRB) 131104
show that this burst may be associated with a gamma-ray transient entitled
Swift~J0644.5-5111. Follow-up observations for potential X-ray and radio counterparts
of FRB~131104/Swift~J0644.5-5111 got null results and provided the upper limits of
the emission flux at 5.5 GHz, 7.5 GHz, $U$-band, and X-ray band.
By assuming this association and
using these upper limits, environmental properties (the fraction of energy
in a magnetic field $\varepsilon_{\rm B}$ and the number density $n$)
of the progenitor system of FRB~131104/Swift~J0644.5-5111 were constrained in the context of
the standard afterglow model that neglects the non-relativistic effect and jet effect
by several groups. In this paper, we adopt a unified afterglow model that takes
into account the non-relativistic effect and jet effect and use the upper limits
of four bands (5.5 GHz, 7.5 GHz, $U$-band, and X-ray) to obtain more stringent
constraints on the parameter space spanned by $\varepsilon_{\rm B}$
and $n$. We thus suggest that
FRB~131104/Swift~J0644.5-5111 might originate from a black hole-neutron star
merger event. Moreover, we calculate multi-band emissions from
a kilonova powered by the radioactivity of $r$-process elements synthesized
in the ejected neutron-rich material and find that the $U$-band emission from the
putative kilonova is significantly lower than the upper limit of the observations.

\end{abstract}

\keywords{gamma-ray burst: general -- radio continuum: general}

\affil{\altaffilmark{1}School of Astronomy and Space Science, Nanjing
University, Nanjing 210093, China; dzg@nju.edu.cn}
\affil{\altaffilmark{2}Key Laboratory of Modern Astronomy and Astrophysics (Nanjing
University), Ministry of Education, China}
\affil{\altaffilmark{3}Department
of Astronomy, University of California, Berkeley, CA 94720-3411, USA}
\affil{\altaffilmark{4}Department of Physics and Astronomy,
University of Nevada, Las Vegas, NV 89154, USA}
\affil{\altaffilmark{5}Department of Physics, Kansas State University,
116 Cardwell Hall, Manhattan, KS 66506, USA}

\section{Introduction}

\label{sec:Intro}

Fast radio bursts (FRBs) discovered one decade ago are radio transients
whose duration timescale is only a few milliseconds~\citep{Lor2007,Rav2016,Kul2014}.
It seems that the event rate of FRBs is very low since only 34 FRBs
have been discovered.\footnote{http://www.astronomy.swin.edu.au/pulsar/frbcat/}
Considering the facts that radio telescopes observe a small area of sky
\footnote{The total instantaneous field-of-view (FoV) of the Molonglo Observatory
Synthesis Telescope (UTMOST) is $\sim 8$ $\rm deg^{2}$
(http://astronomy.swin.edu.au/research/utmost/?page\_id=32);
the total instantaneous FoV of the Five-hundred-meter Aperture
Spherical radio Telescope (FAST) is 0.15 $\rm deg^{2}$ \citep{Li2017},
and based on the method provided by \citet{Li2017}, the computed
total instantaneous FoV of the Parkes radio telescope is 2.24 $\rm deg^{2}$.} and
that the observed band is rather narrow, however, the actual explosion rate of FRBs must be
rather high, about $2.8\times10^{3}$ Gpc$^{-3}$ yr$^{-1}$,
and about $10^{4}$ bursts per day all sky~\citep{Has2013}.
These FRBs may be divided into two subclasses, repeating and non-repeating FRBs.
Except for FRB~121102 which is a repeating FRB that has been recorded
more than 140 bursts \citep[cf.][]{Pal2018},\footnote{Most of the bursts 
have not been published to date, see,
Spitler, the 2017 Aspen FRB conference (http://aspen17.phys.wvu.edu/Spitler.pdf).}
and the other are non-repeating.\footnote{\citet{Pir2017}
proposed that FRB~110220 and FRB~140514 might be the two bursts
originated a repeating FRB since their positions are close to each other.}

The typical frequency and the observed peak flux of FRBs are $\sim1$ GHz and $\sim1$ Jy,
respectively.\footnote{However, the peak fluxes of some FRBs are significantly higher than 1 Jy, e.g.,
the peak flux of FRB~150807 is $128 \pm 5$~Jy \citep{Rav2016}.} Based on these facts,
the inferred brightness temperature can be up to $\sim~10^{37}$ K, which is most likely
to be reproduced by the coherent radiation mechanism~\citep{Lor2007,Kat2014,Yang2017}.
The dispersion measure (DM) of a FRB can be calculated from the
difference of the arrival time between different frequencies.
By comparing the DMs of the FRBs and that of our Galaxy at the same directions,
one can find that the former are significantly larger than the latter~\citep{Tho2013}.
Research for the distributions of observed FRBs indicate that the positions of these FRBs
are random rather than concentrating to the disk of the Galaxy~\citep{Cal2017},
indicating that the FRBs have a cosmological origin rather than a Galactic origin.

In the past decade, many theoretical models have been proposed to explain FRBs,
including magnetar giant flares \citep{Kul2014},
giant pulses from young pulsars \citep{Con2016},
planetary companion around pulsars \citep{Mot2014},
neutron star (NS)-black hole (BH) transitions \citep{Fal2014},
NS-white dwarf (WD) mergers \citep{Gu2016},
NS-NS mergers \citep{Aba2010,Wang2016},
WD-WD mergers \citep{Kas2013},
NS-BH mergers \citep{Min2015},
pulsar-asteroid collisions \citep{Geng2015,Dai2016b},
cosmic string collisions \citep{Zad2015},
charged BH-BH mergers \citep{Zhang2016},
light sails of extragalactic civilizations \citep{Lin2017}, and so on.

Searching and observing the electromagnetic counterparts at other bands,
e.g., X-ray, ultraviolet (UV), optical, as well as infrared (IR), are
helpful for determining the nature of FRBs. However, discovering these
electromagnetic counterparts of FRBs are very difficult since the
positional accuracy for FRBs are very poor so that their host galaxies
can hardly be fixed. Furthermore, a major of FRBs were found from
historical records and the electromagnetic counterparts at other bands
cannot be confirmed.

Nevertheless, two (FRB~150418 and FRB~131104) of 33 non-repeated FRBs
might have electromagnetic counterpart candidates at other bands observed.
\citet{Kea2016} declared that they found bright radio transients on 5.5 GHz
and 7.5 GHz from 2 hours to 6 days after with FRB~150418.
The results of FRB~150418 are controversial since the observations of
the Very Large Array (VLA) suggested that the 5.5 GHz and 7.5 GHz radio
emission detected by the Australia Telescope Compact Array (ATCA) 2 hours
to 6 days after the trigger of FRB~150418 might be generated by the activity
of an active galactic nucleus (AGN) rather than FRB~150418 \citep{Wil2016}.

A gamma-ray transient, Swift~J0644.5-5111, has been reported to be associated
with FRB~131104, the confidence level is 3.2$\sigma$~\citep{DeL2016}.
The duration $T_{90}$ and the fluence $S_{\gamma}$ of Swift~J0644.5-5111
are $\approx 377$ s and $\approx 4\times10^{-6}$ erg cm$^{-2}$, respectively;
the DM of FRB~131104 is $779 \pm 1$ pc cm$^{-3}$. The redshift ($z$) of
FRB~131104/Swift~J0644.5-5111 inferred from the DM
is $\approx 0.55$.\footnote{The inferred redshift
relies on the assumption that the DM is dominated by the intergalactic medium.
A large host contribution would result in a smaller redshift.}
Therefore, the isotropic gamma-ray energy $E_{\gamma,{\rm iso}}$ of
Swift~J0644.5-5111 is $\approx 5\times10^{51}$~erg~\citep{DeL2016}. The follow-up
observations performed by the XRT and the UVOT on board {\em Swift}
two days after {\em Swift} was triggered by Swift~J0644.5-5111 didn't
detect X-ray and optical counterparts~\citep{DeL2016}. \citet{Sha2017}
observed the 5.5 GHz and 7.5 GHz counterparts of FRB~131104 three days to
2.5 years after the FRB trigger and found that the upper limits of the
5.5 GHz and 7.5 GHz are $70$ $\mu $Jy and $100$ $\mu$Jy, respectively.

A powerful tool to determine the progenitor system of Swift~J0644.5-5111 is to
research the properties of its environment, e.g., the number density
($n$) of the interstellar medium (ISM) surrounding the progenitor.
Assuming that Swift~J0644.5-5111 was associated with FRB~131104 and
using the upper limit of 5.5 GHz counterpart of FRB~131104 reported
by \citet{Sha2017}, \citet{Mur2017} constrained the density of the
ISM of FRB~131104 with the external shock model of afterglows and
concluded that $n\lesssim 2\times 10^{-3}$~cm~$^{-3}$.
By using the same model and the upper limits of radio counterparts and
adopting different values of redshift, the fraction of energy in
magnetic field ($\varepsilon_{\rm B}$), and the fraction of energy
in electrons ($\varepsilon_{\rm e}$), \citet{Gao2017} constrained
the parameter space of the ISM density. According to the detected
limit obtained by the XRT and employing the afterglow model,
\citet{Dai2016a} got a stringent constraint,
$n\lesssim 2.6\times 10^{-4}$~cm$^{-3}$.

While some research mentioned above \citep{Mur2017,Gao2017,Dai2016a}
well constrained the density of the ISM of Swift~J0644.5-5111/FRB~131104,
they assumed that the explosion producing the gamma-ray burst (GRB) is
isotropic and relativistic, and didn't take into account the jet effect
and the evolution from ultra-relativistic to non-relativistic. However,
these two effects cannot be neglected since the outflows launched by the
explosions of a massive star or the merger of NS-NS/BH binaries are jets
rather than isotropic fireballs, and the ISM would decelerate the relativistic
jet and the jet would eventually become non-relativistic if the jet
sweeps up an enough mass of the ISM.

In this paper, we also suppose that both FRB~131104 and
Swift~J0644.5-5111 originate from an explosion of a massive star or
a merging of NS-NS/BH binary and adopt the unified dynamic afterglow
model proposed by \citet{Huang1999,Huang2000a,Huang2000b} which describes ultra-relativistic,
trans-relativistic, and non-relativistic outflows to pose more stringent
constraints on the density of the ISM and other parameters of
Swift~J0644.5-5111/FRB~131104. The jet effect and non-relativistic
effect would be taken into account throughout this paper.
The remainder of this paper is organized as follows.
In Section \ref{sec:model} we present our adopted model. In Section
\ref{sec:fit}, we constrain the parameters. Our discussion and conclusions
are presented in Sections \ref{sec:dis} and \ref{sec:con}, respectively.

\section{The model}

\label{sec:model}

We assume that FRB~131104 is associated with Swift~J0644.5-5111 and originates
from the core collapse of a massive star or the merger of binary compact stars.
It is suggested that a relativistic outflow could be generated during these
processes. The relativistic outflow whose velocity is significantly larger than
the sound velocity of the ISM sweeps up this external medium, producing
a forward shock which accelerates the electrons. The accelerated electrons
emit photons which constitute multi-band afterglows~\citep{Sar1998}.
In order to accurately describe the outflow evolution from ultra-relativistic to
non-relativistic stages, \citet{Huang1999,Huang2000a,Huang2000b}
developed a unified dynamic model for an afterglow which can be applied
both the completely adiabatic case and the completely radiative case.
The equation of the dynamic evolution is given by %
\begin{equation}
\frac{d\gamma }{dm}=-\frac{\gamma^{2}-1}{M_{\rm ej}+\epsilon
m+2(1-\epsilon )\gamma m},
\end{equation}%
where $\gamma$ is the Lorentz factor of the shock wave, $m$ is the mass
of the ISM swept up by the shock, and $M_{\rm ej}$ and $\epsilon$ are
the outflow mass and the radiative efficiency respectively.
To solve this equation, three additional auxiliary equations are
required by
\begin{equation}
\frac{dR}{dt}=\beta c\gamma (\gamma +\sqrt{\gamma^{2}-1}),
\end{equation}
\begin{equation}
\frac{dm}{dR}=2\pi R^{2}(1-\cos \theta )nm_{\rm p},
\end{equation}%
\begin{equation}
\frac{d\theta }{dt}=\frac{c_{s}(\gamma +\sqrt{\gamma ^{2}-1})}{R},
\end{equation}
where $\theta$ is the half-angle of the outflow, $c_{s}$ is the sound velocity
of the outflow, $R$ is the radius of the shock, and $t$ is the observed time.
The dynamic evolution of the shock can be obtained by solving these equations.

According to the standard afterglow model, the distribution of the electrons
accelerated by the forward shock is a power-law function of
$\gamma_{\rm e}$ \citep{Mes1997,Sar1998}:
$N_{\rm e}(\gamma _{\rm e}){\rm d}\gamma_{\rm e}\propto\gamma_{\rm e}^{-p}{\rm d}\gamma_{\rm e}$,
where $p$ is the spectral index and is fixed to be 2.5 here. The minimum
Lorentz factor of the accelerated electrons is
\begin{equation}
\gamma_{\rm m}=\frac{p-2}{p-1}\frac{m_{\rm p}}{m_{\rm e}}\varepsilon
_{\rm e}(\gamma -1)+1,
\end{equation}
where $m_{\rm p}$ and $m_{\rm e}$ are the masses of the protons and electrons,
respectively. $\varepsilon_{\rm e}$ is the fraction of the shock energy going
into the electrons. The critical Lorentz factor above which the relativistic
electrons would emit synchrotron radiation in the dynamical time $t$ is%
\begin{equation}
\gamma _{\rm c}=(1+z)\frac{6\pi m_{\rm e}c}{\sigma _{\rm T}\gamma B^{2}t},
\end{equation}
where $c$ is the speed of light in a vacuum, $\sigma_{\rm T}$ is the Thomson
scattering cross-section, $B=(32{\pi}m_{\rm p}\varepsilon_{\rm B}n)^{\frac{1}{2}}\gamma c$
is the magnetic field behind the shock and $\varepsilon_{\rm B}$ is the fraction of
energy in a magnetic field. Hence, the two characteristic frequencies of the
synchrotron radiation are%
\begin{equation}
\nu_{\rm m}=3.3\times 10^{11}(1+z)^{\frac{1}{2}}\varepsilon
_{\rm e,-1}^{2}\varepsilon_{\rm B,-2}^{\frac{1}{2}}E_{51.7}^{\frac{1}{2}}t_{\text{%
day}}^{-\frac{3}{2}}~\text{Hz},
\end{equation}
and
\begin{equation}
\nu _{c}=3.8\times 10^{17}(1+z)^{-\frac{1}{2}}\varepsilon _{B,-2}^{-\frac{3}{%
2}}E_{51.7}^{-\frac{1}{2}}n_{-2}^{-1}t_{\rm day}^{-\frac{1}{2}}~{\rm Hz}.
\end{equation}

Assuming that $\varepsilon_{\rm e} = 0.1$
and $\varepsilon_{\rm B} = 0.01$, $\nu _{m}$ becomes smaller
than $\nu_{c}$ very quickly. The electrons would be in the slow-cooling phase
about two days after the explosion. Considering the synchrotron self-absorption effect,
the synchrotron self-absorption frequency at the radio band is %
\begin{eqnarray}
\nu _{a}=\left\{
\begin{array}{l}
2.4\times 10^{8}(1+z)^{-1}\varepsilon _{e,-1}^{-1}\varepsilon _{B,-2}^{\frac{%
1}{5}}E_{51.7}^{\frac{1}{5}}n_{-2}^{\frac{3}{5}}\text{ Hz}, \\
~~~~~~~~~~~~~\text{ \ \ \ \ \
\ \ \ \ \ \ \ \ \ \ \ \ \ \ \ \ \ }\nu _{a}<\nu _{m}<\nu _{c}; \\
8.6\times 10^{9}(1+z)^{\frac{p-6}{2(p+4)}}\varepsilon _{e,-1}^{\frac{2(p-1)}{%
p+4}}\varepsilon _{B,-2}^{\frac{p+2}{2(p+4)}}E_{51.7}^{\frac{p+2}{2(p+4)}%
}\\{\times} n_{-2}^{\frac{2}{p+4}}t_{\rm day}^{-\frac{3p+2}{2(p+4)}}\text{ Hz},~~~~~~~~\text{ \ \ }%
\nu _{m}<\nu _{a}<\nu _{c};%
\end{array}%
\right.
\end{eqnarray}
the observed flux is %
\begin{equation}
F_{\nu }=\left\{
\begin{array}{l}
(\frac{\nu }{\nu _{a}})^{2}(\frac{\nu _{a}}{\nu _{m}})^{\frac{1}{3}}F_{\nu
,\max },\text{ \ \ \ \ \ \ \ \ \ }\nu <\nu _{a}; \\
(\frac{\nu }{\nu _{m}})^{\frac{1}{3}}F_{\nu ,\max },\text{ \ \ \ \ \ \ \ \ \
\ \ \ \ \ \ \ }\nu _{a}<\nu <\nu _{m}; \\
(\frac{\nu }{\nu _{m}})^{-\frac{p-1}{2}}F_{\nu ,\max },\text{ \ \ \ \ \ \ \
\ \ \ \ }\nu _{m}<\nu <\nu _{c}; \\
(\frac{\nu }{\nu _{c}})^{-\frac{p}{2}}(\frac{\nu _{c}}{\nu _{m}})^{-\frac{p-1%
}{2}}F_{\nu ,\max },\text{ \ \ }\nu _{c}<\nu <\nu _{\max };%
\end{array}%
\right. \text{\ }
\end{equation}%
when $\nu _{a}<\nu _{m}<\nu _{c}$, and
\begin{equation}
F_{\nu }=\left\{
\begin{array}{l}
(\frac{\nu }{\nu _{m}})^{2}(\frac{\nu _{m}}{\nu _{a}})^{\frac{p+4}{2}}F_{\nu
,\max },\text{ \ \ \ \ \ \ }\nu <\nu _{m}; \\
(\frac{\nu }{\nu _{a}})^{\frac{5}{2}}(\frac{\nu _{a}}{\nu _{m}})^{-\frac{p-1%
}{2}}F_{\nu ,\max },\text{ \ \ \ \ }\nu _{m}<\nu <\nu _{a}; \\
(\frac{\nu }{\nu _{m}})^{-\frac{p-1}{2}}F_{\nu ,\max },\text{ \ \ \ \ \ \ \
\ \ \ \ }\nu _{a}<\nu <\nu _{c}; \\
(\frac{\nu }{\nu _{c}})^{-\frac{p}{2}}(\frac{\nu _{c}}{\nu _{m}})^{-\frac{p-1%
}{2}}F_{\nu ,\max },\text{ \ \ }\nu _{c}<\nu <\nu _{\max };%
\end{array}%
\right.
\end{equation}%
when $\nu _{m}<\nu _{a}<\nu _{c}$. Here, $\nu _{\max}$ is the characteristic frequency
of the electron having the largest Lorentz factor $\gamma _{\max}
\simeq [6\pi q_{\rm e}/(\sigma_{\rm T}B)]^{1/2}$.
$F_{\nu ,\max }$ is the peak value of the observed flux %
\begin{eqnarray}
F_{\nu ,\max }&=&(1+z)\frac{N_{e,\text{tot}}P_{\nu ,\max }}{4\pi D_{L}^{2}}\nonumber \\ %
&=&5.7\times 10^{2}(1+z)\varepsilon _{B,-2}^{\frac{1}{2}}E_{51.7}n_{-2}^{\frac{%
1}{2}}D_{L,28}^{-2}\text{}~\mu\text{Jy},\nonumber \\
\end{eqnarray}%
where $D_L$ is the luminosity distance to the source.

Some relevant works (e.g., \citealt{Mur2017,Gao2017,Dai2016a}) in the literature
do not take into account the non-relativistic effect, the jet effect and the lateral
expansion effect. However, all these effects are included in our numerical calculations.

\section{Constraining the parameters}
\label{sec:fit}

In this section, we use the equations of the unified dynamic afterglow model
listed in section \ref{sec:model} as well as the upper limits of the
observations of radio (5.5 GHz and 7.5 GHz), $U$-band, and
X-ray counterparts to constrain the ISM density ($n$) and $\varepsilon_{\rm B}$.
Some parameters are fixed: $z=0.55$, $E_{\gamma,{\rm iso}}=5\times 10^{51}$ erg,
$\theta = 0.1$, $p = 2.5$, $\varepsilon_{\rm e} = 0.1$.
Thus, the light curves depend on the values of $\varepsilon_{\rm B}$ and $n$
which are the parameters needed to be constrained. For comparison, we also calculate the multi-band
afterglows from both for the isotropic outflow and for the jet
whose half-angle is supposed to be 0.1 here.
\citet{San2014} presented a systematic study on magnetic fields in GRB external forward shocks
and found that the range and the median value of $\varepsilon_{\rm B}$ are $10^{-4}-10^{-0.5}$
and $\approx 10^{-2}$, respectively. Hence, we set $\varepsilon_{\rm B}$ = 0.0001, 0.001,
0.01, and 0.1.

\subsection{The parameters inferred from the models}

\begin{figure}[tbph]
\begin{center}
\includegraphics[width=0.35\textwidth,angle=0]{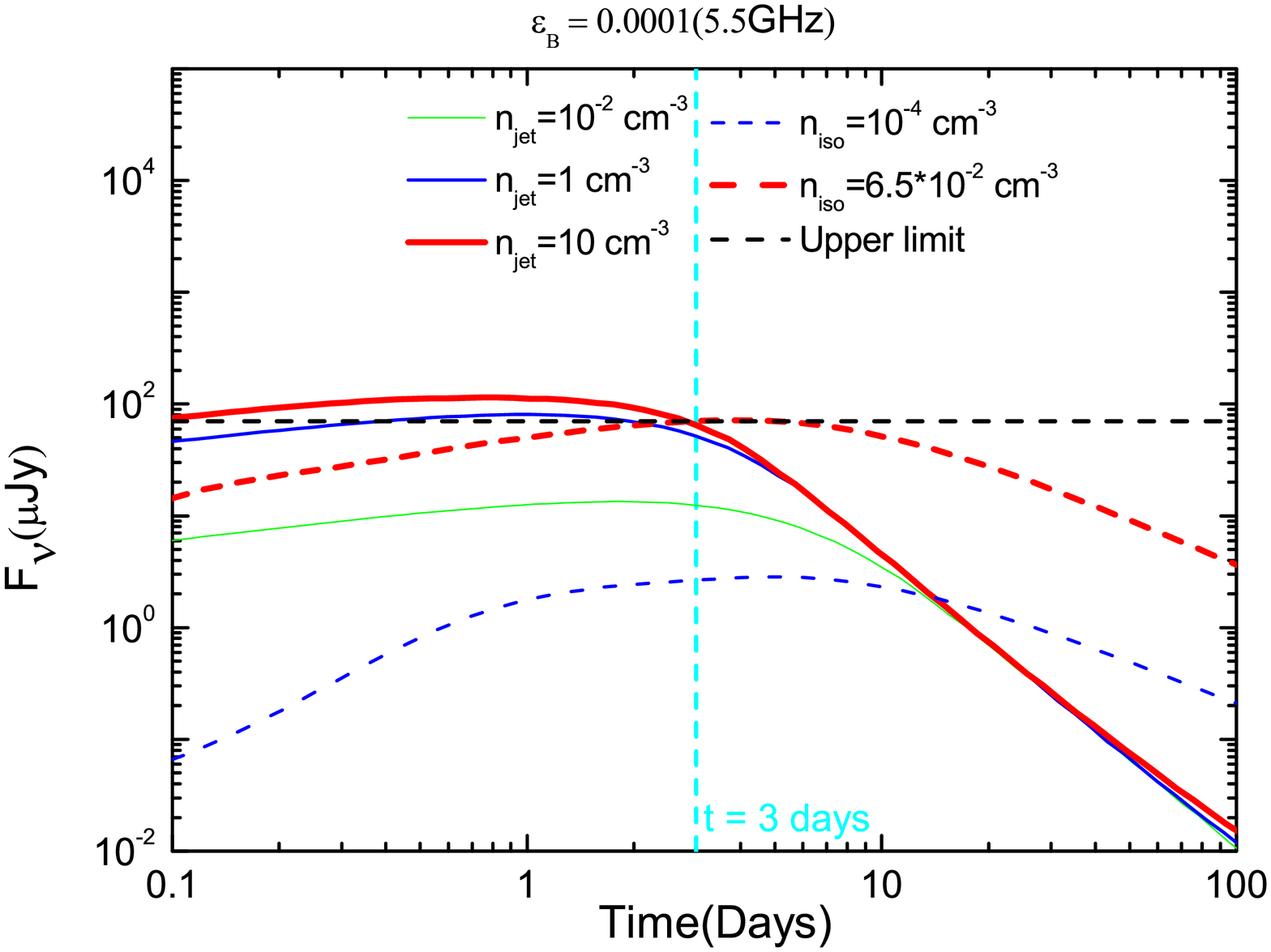} %
\includegraphics[width=0.35\textwidth,angle=0]{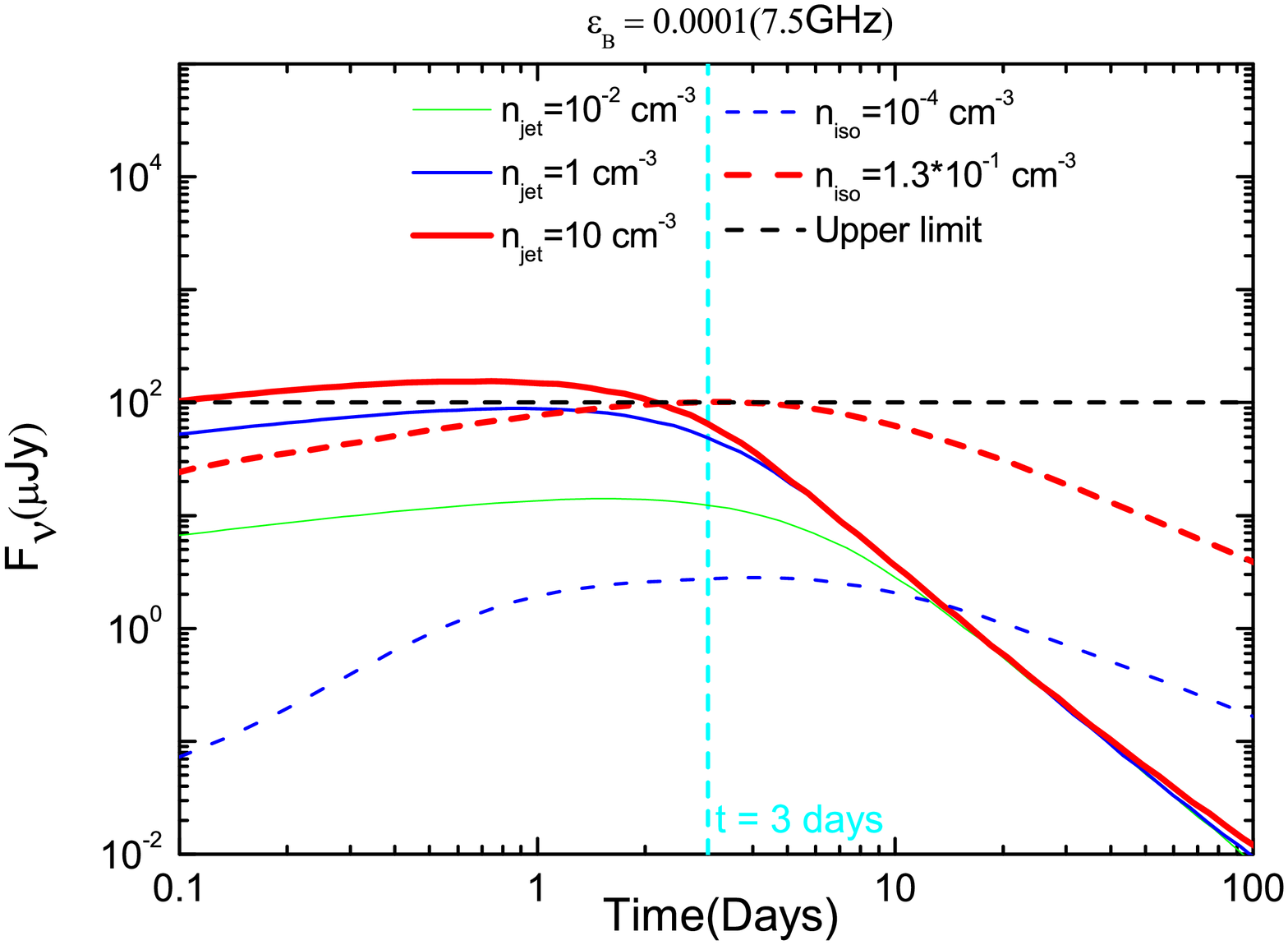}\\[0pt]
\includegraphics[width=0.35\textwidth,angle=0]{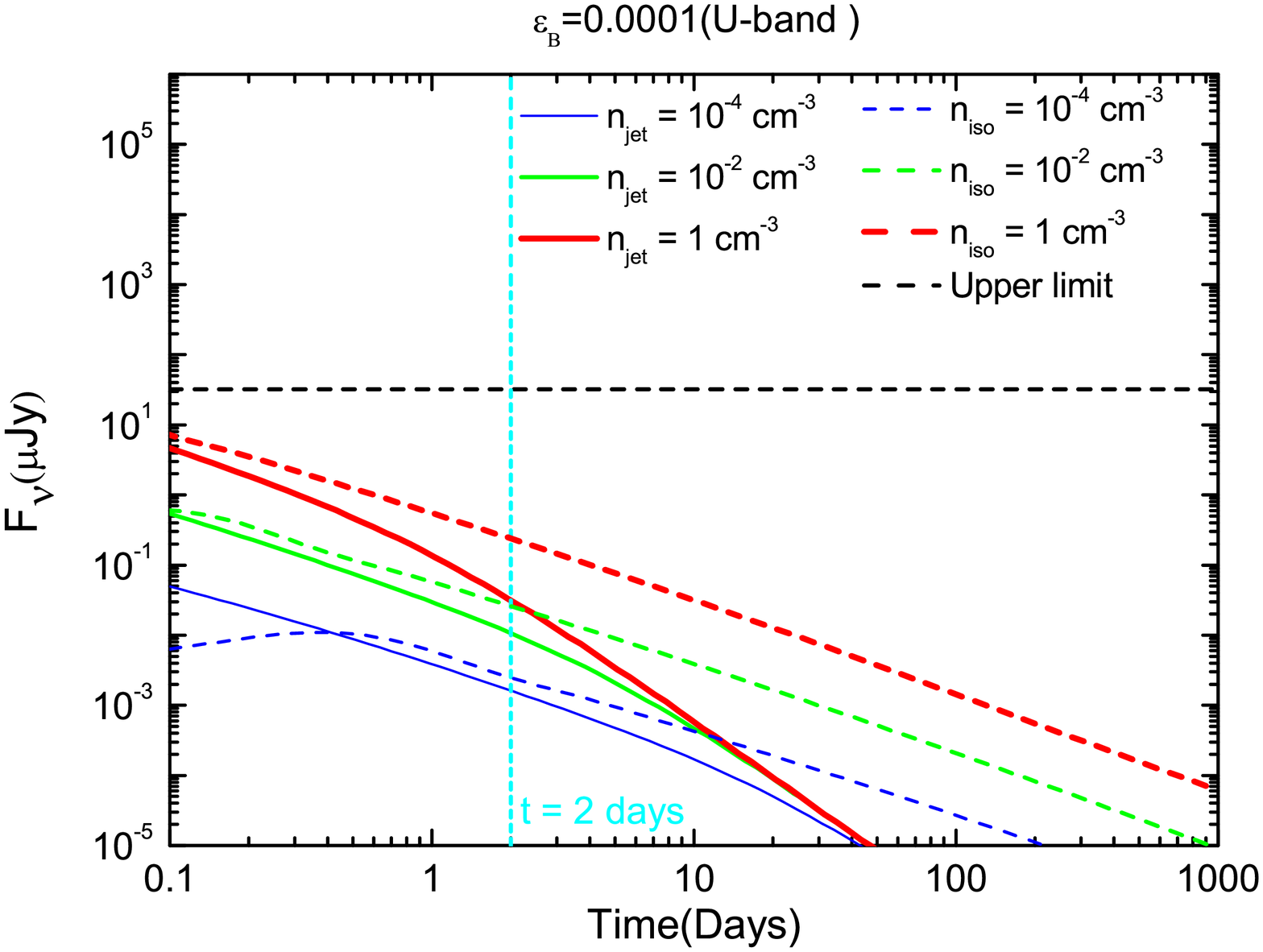} %
\includegraphics[width=0.35\textwidth,angle=0]{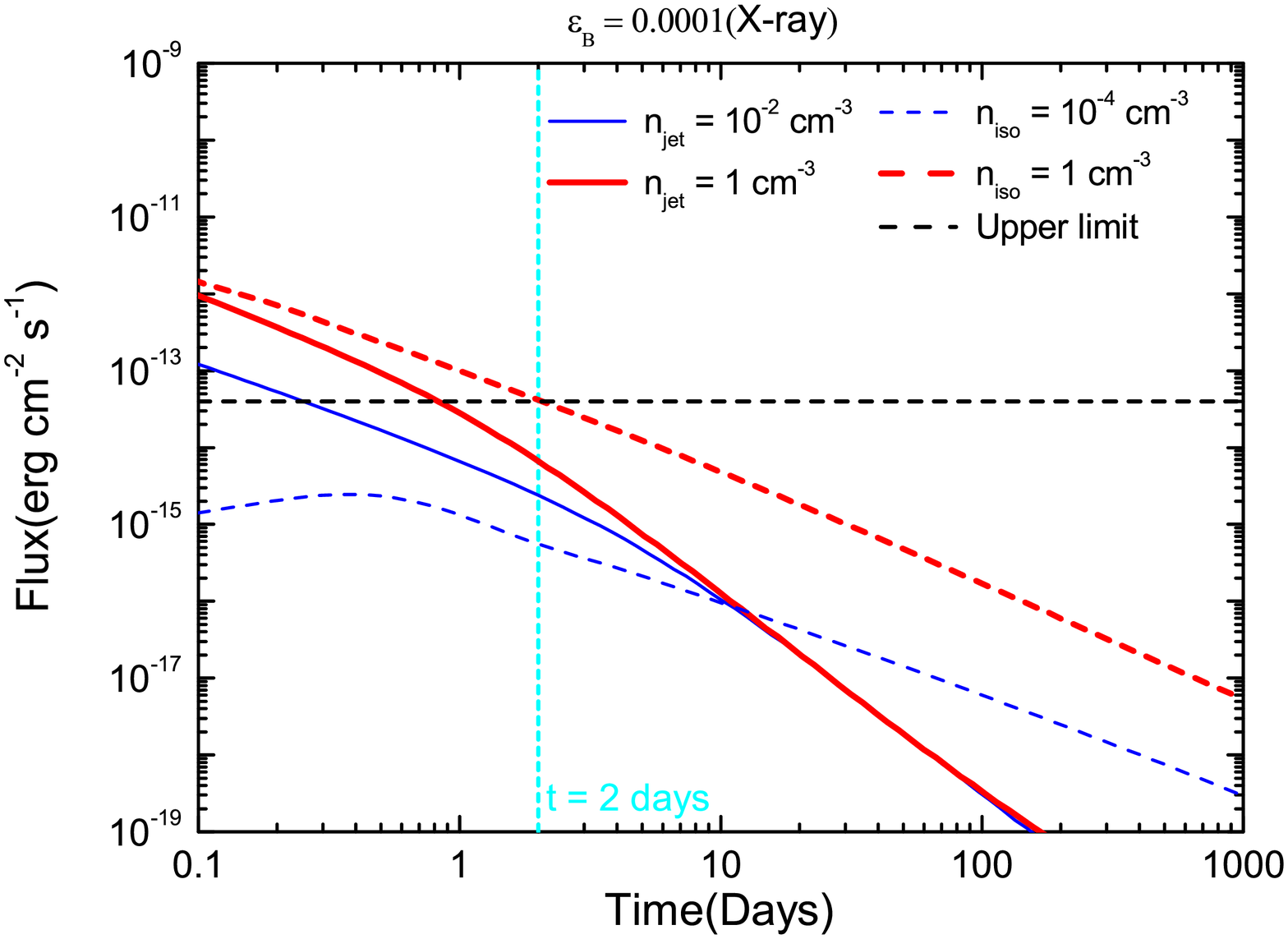} %
\end{center}
\caption{The light curves of 5.5 GHz, 7.5 GHz, U-band, and X-ray when the
fraction of the shock energy going into the electrons $\epsilon_B$
is set to be 0.0001. The solid lines correspond to the jet case while
the dashed lines correspond to the isotropic-explosion case. The black dashed
lines represent the detection limits of each band, while the vertical
cyan lines represent the start times of observations.
The light curves yielded by different densities (indicated in the panels)
are represented by different colors, thicknesses and styles.}
\label{fig:0.0001}
\end{figure}

\begin{figure}[tbph]
\begin{center}
\includegraphics[width=0.35\textwidth,angle=0]{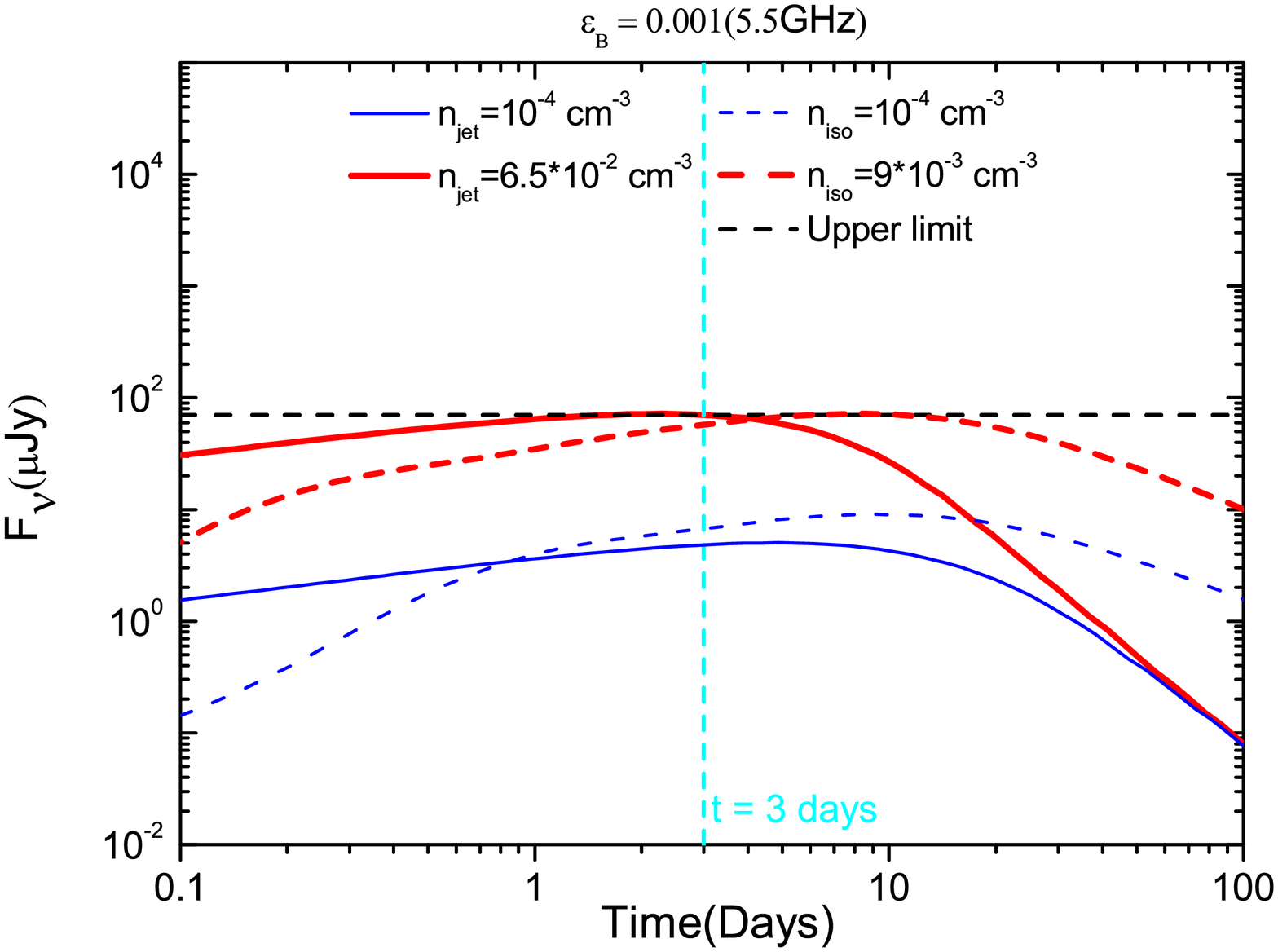} %
\includegraphics[width=0.35\textwidth,angle=0]{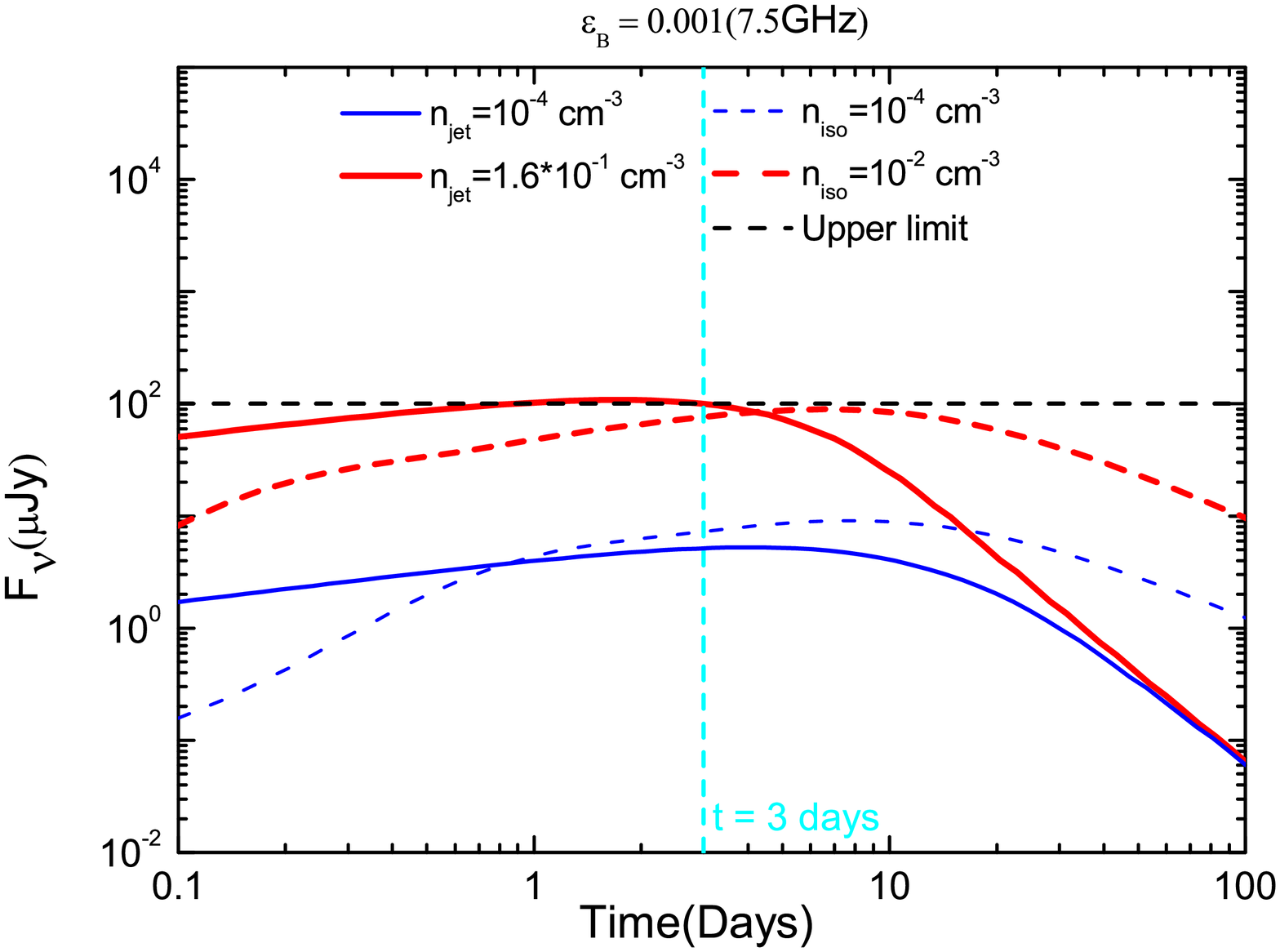}\\[0pt]
\includegraphics[width=0.35\textwidth,angle=0]{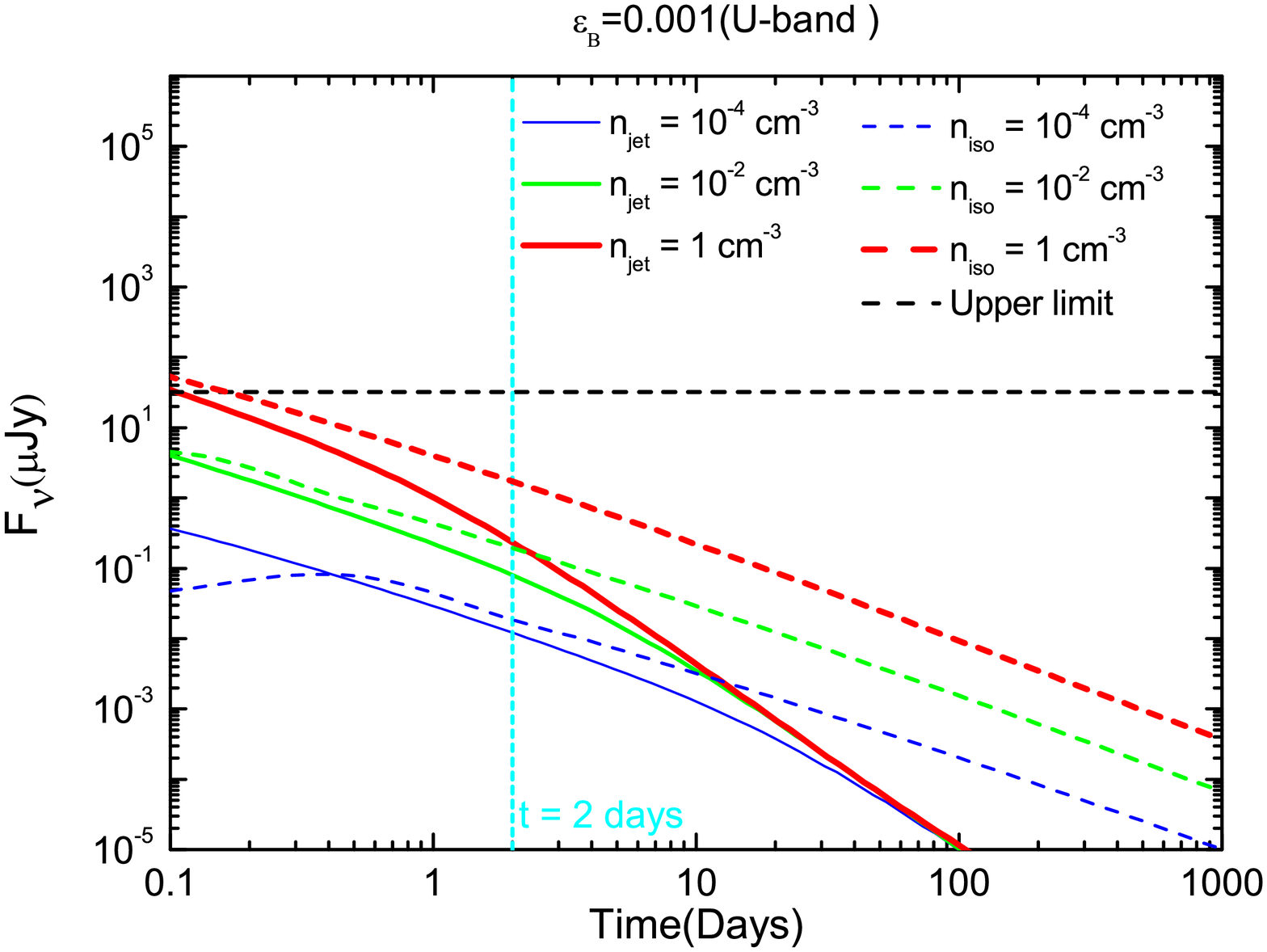} %
\includegraphics[width=0.35\textwidth,angle=0]{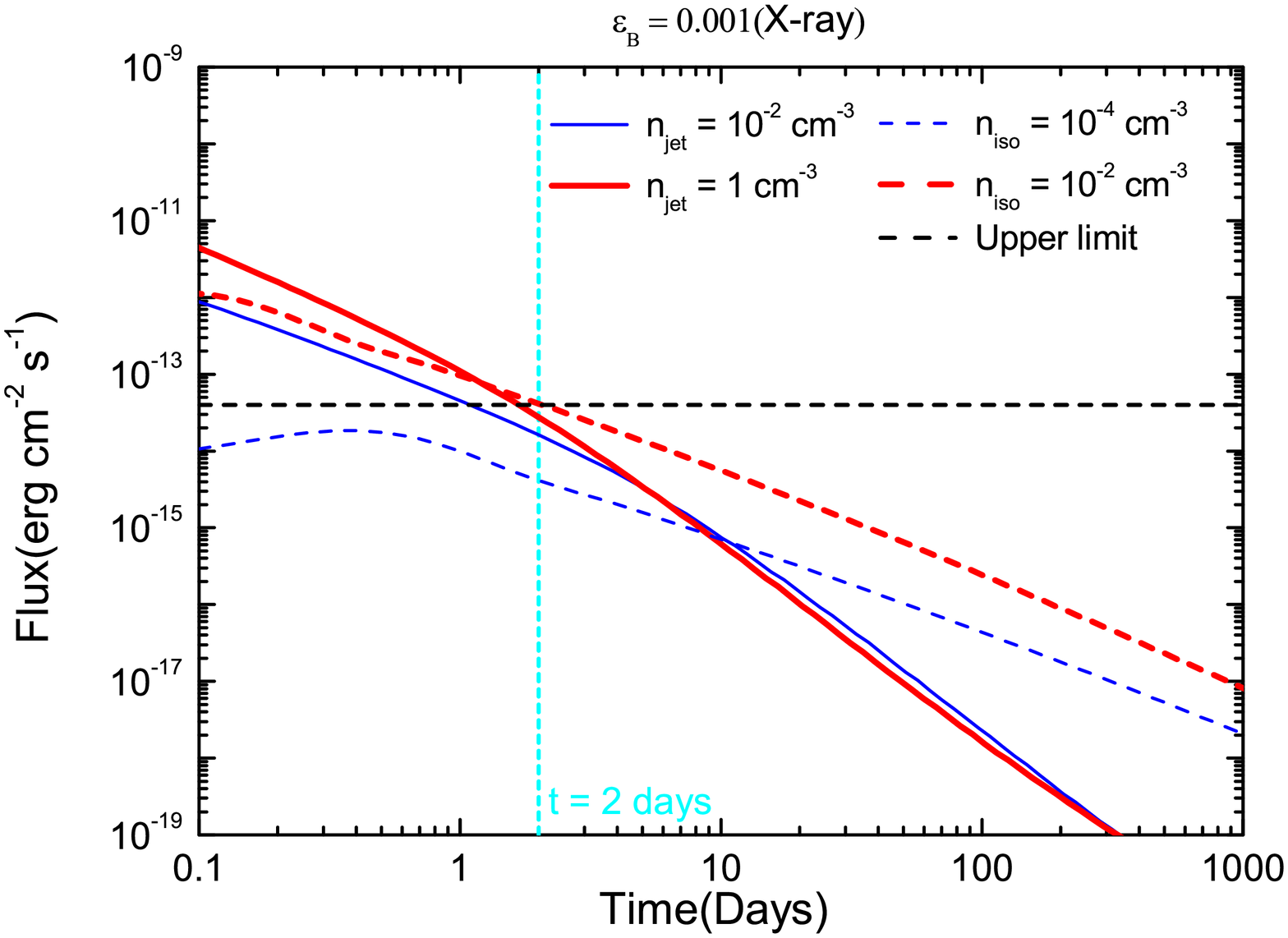} %
\end{center}
\caption{Same as in Fig. \ref{fig:0.0001} but $\epsilon_B$ is 0.001.}
\label{fig:0.001}
\end{figure}

\begin{figure}[tbph]
\begin{center}
\includegraphics[width=0.35\textwidth,angle=0]{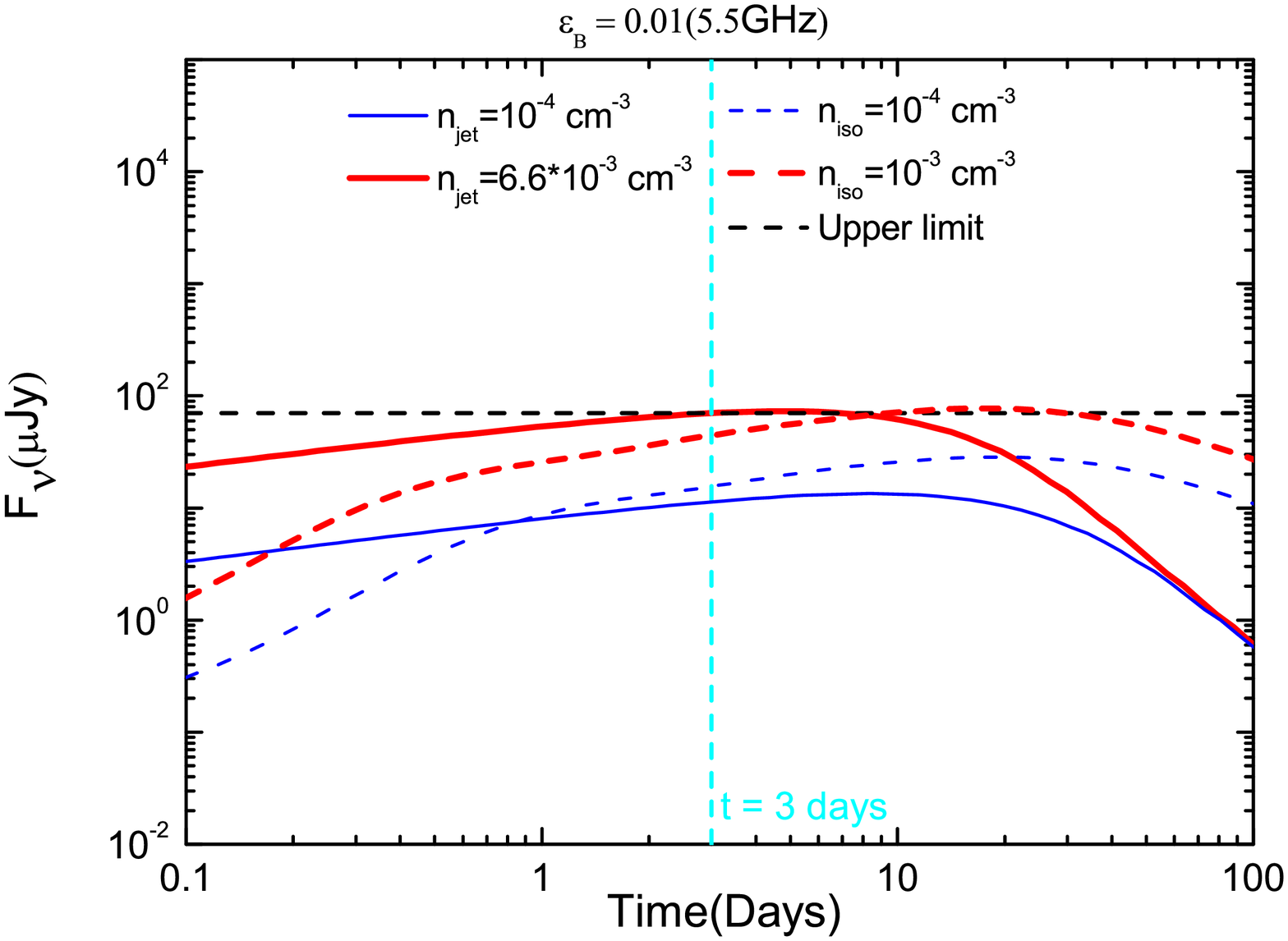} %
\includegraphics[width=0.35\textwidth,angle=0]{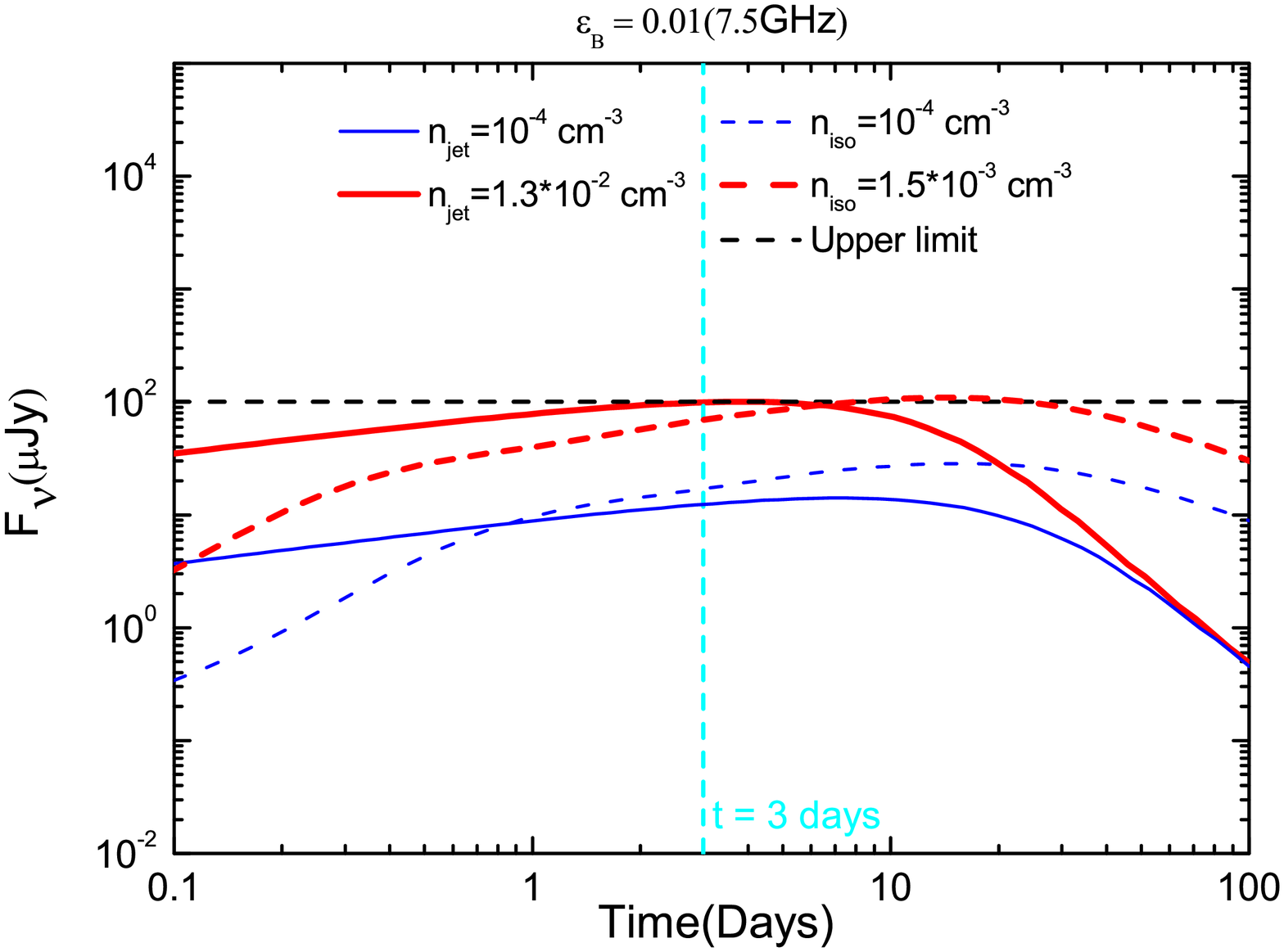}\\[0pt]
\includegraphics[width=0.35\textwidth,angle=0]{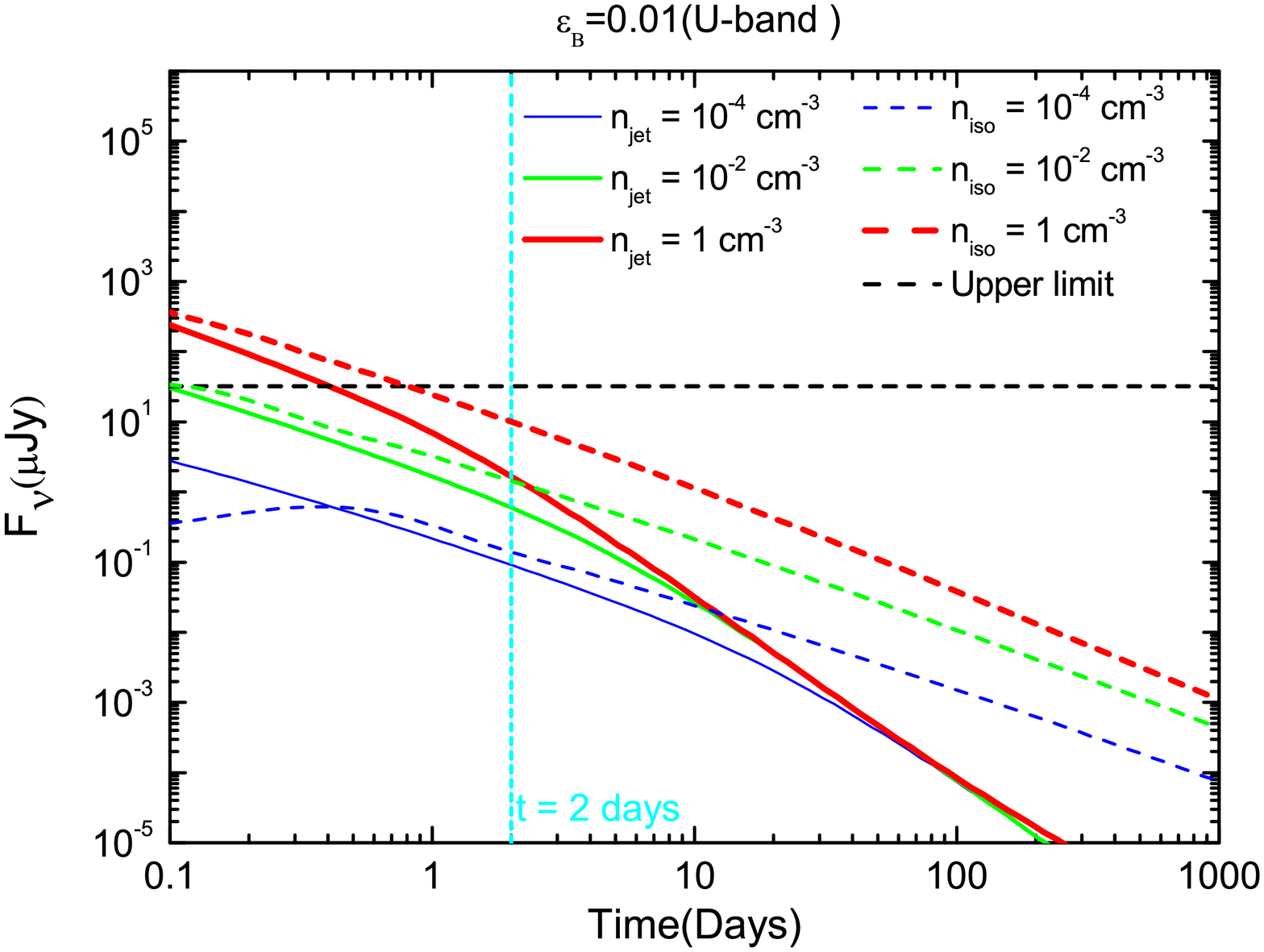} %
\includegraphics[width=0.35\textwidth,angle=0]{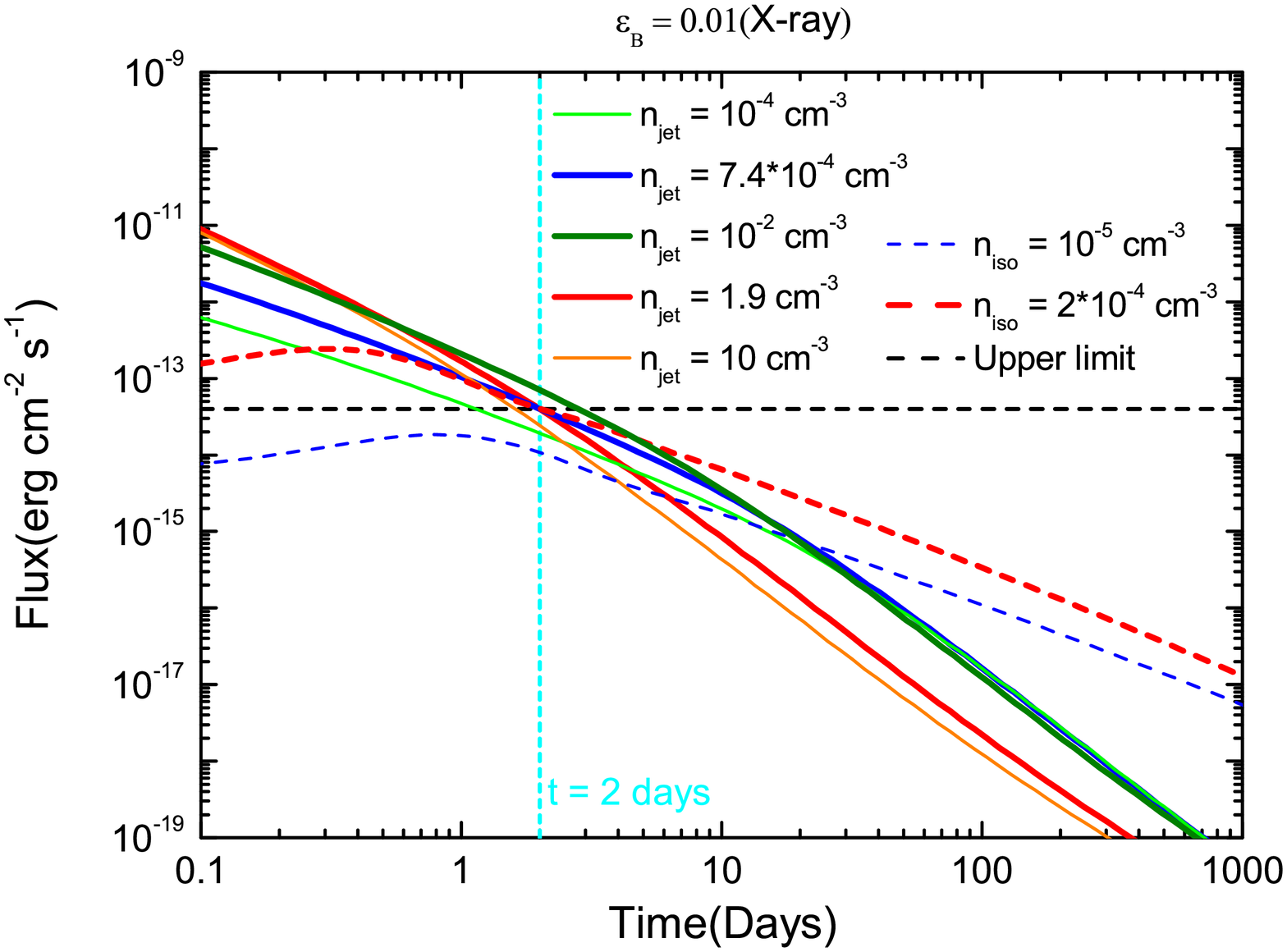} %
\end{center}
\caption{Same as in Fig. \ref{fig:0.0001} but $\epsilon_B$ is 0.01. The lines in the last panel
are more than those of the remaining panels of Fig. \ref{fig:0.0001} and Fig. \ref{fig:0.001} since the lines
reach the upper limit at $t = 2$ days if $n = 1.9$ cm$^{-3}$ or $7.4 \times 10^{-4}$ cm$^{-3}$
and additional light curves corresponding to other three possible cases ($n > 1.9$ cm$^{-3}$,
$7.4\times 10^{-4}$ cm$^{-3}$ $< n <$ 1.9 cm$^{-3}$, and $n < 7.4\times 10^{-4}$ cm$^{-3}$)
are given.}
\label{fig:0.01}
\end{figure}

\begin{figure}[tbph]
\begin{center}
\includegraphics[width=0.35\textwidth,angle=0]{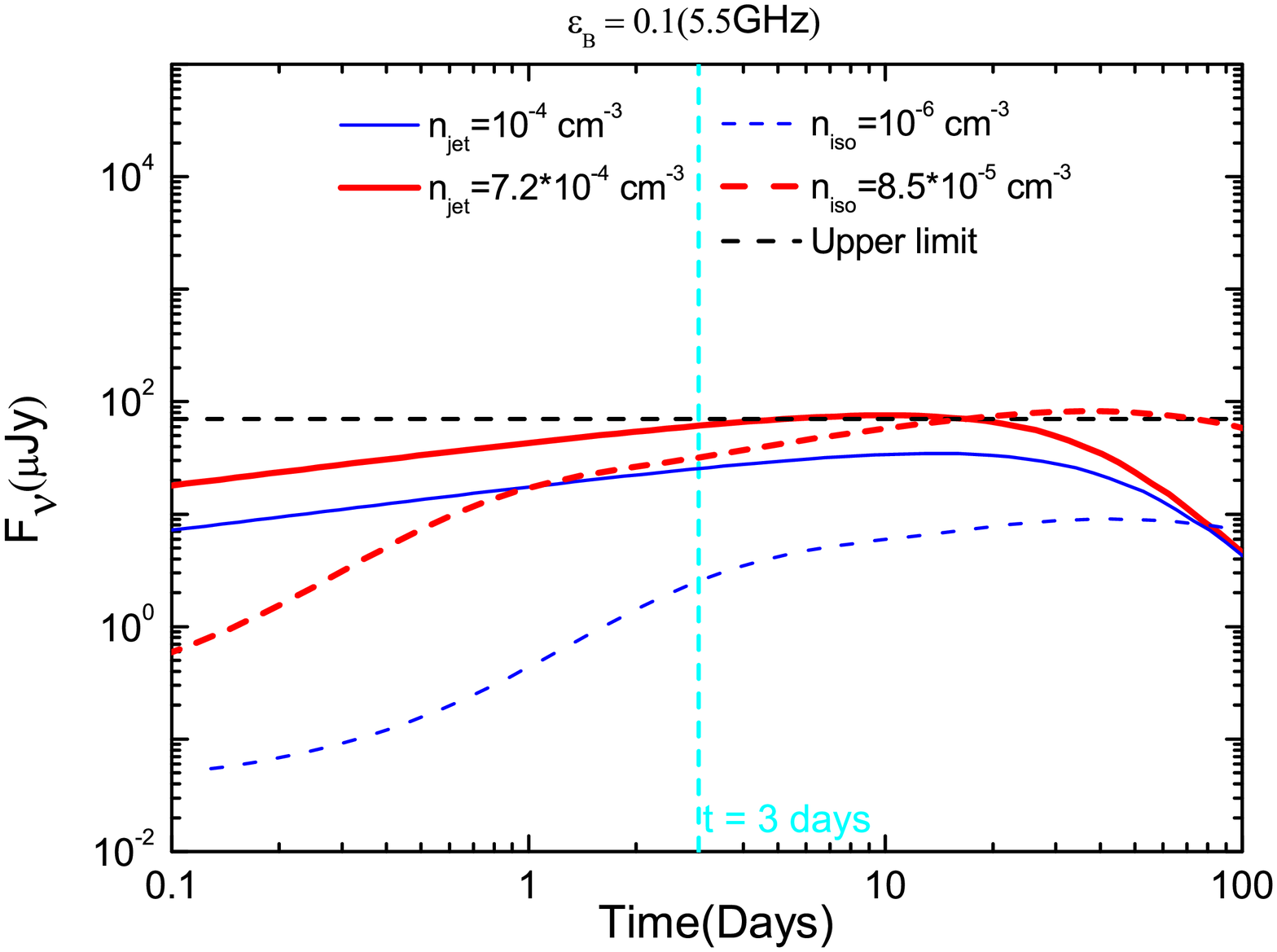} %
\includegraphics[width=0.35\textwidth,angle=0]{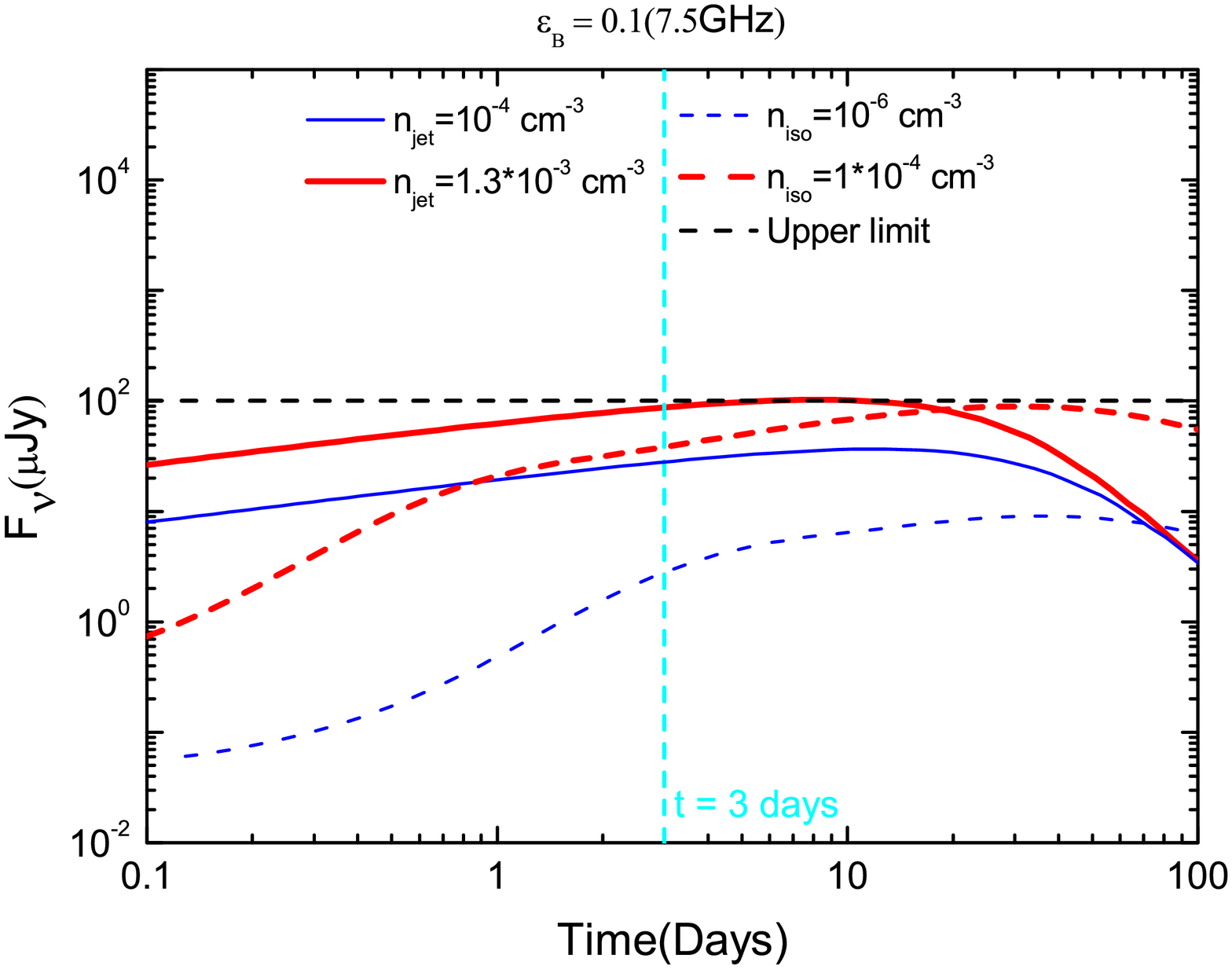}\\[0pt]
\includegraphics[width=0.35\textwidth,angle=0]{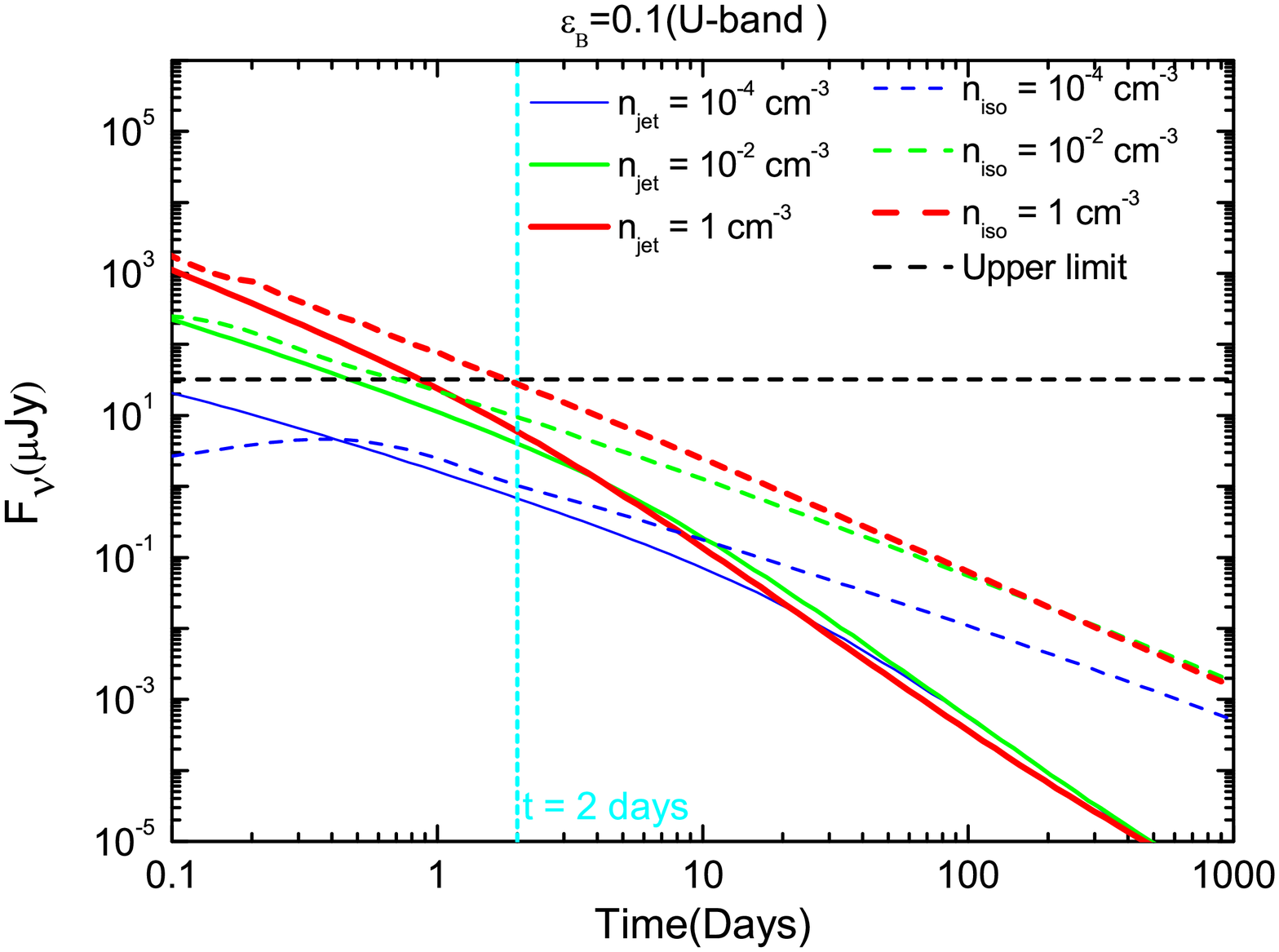} %
\includegraphics[width=0.35\textwidth,angle=0]{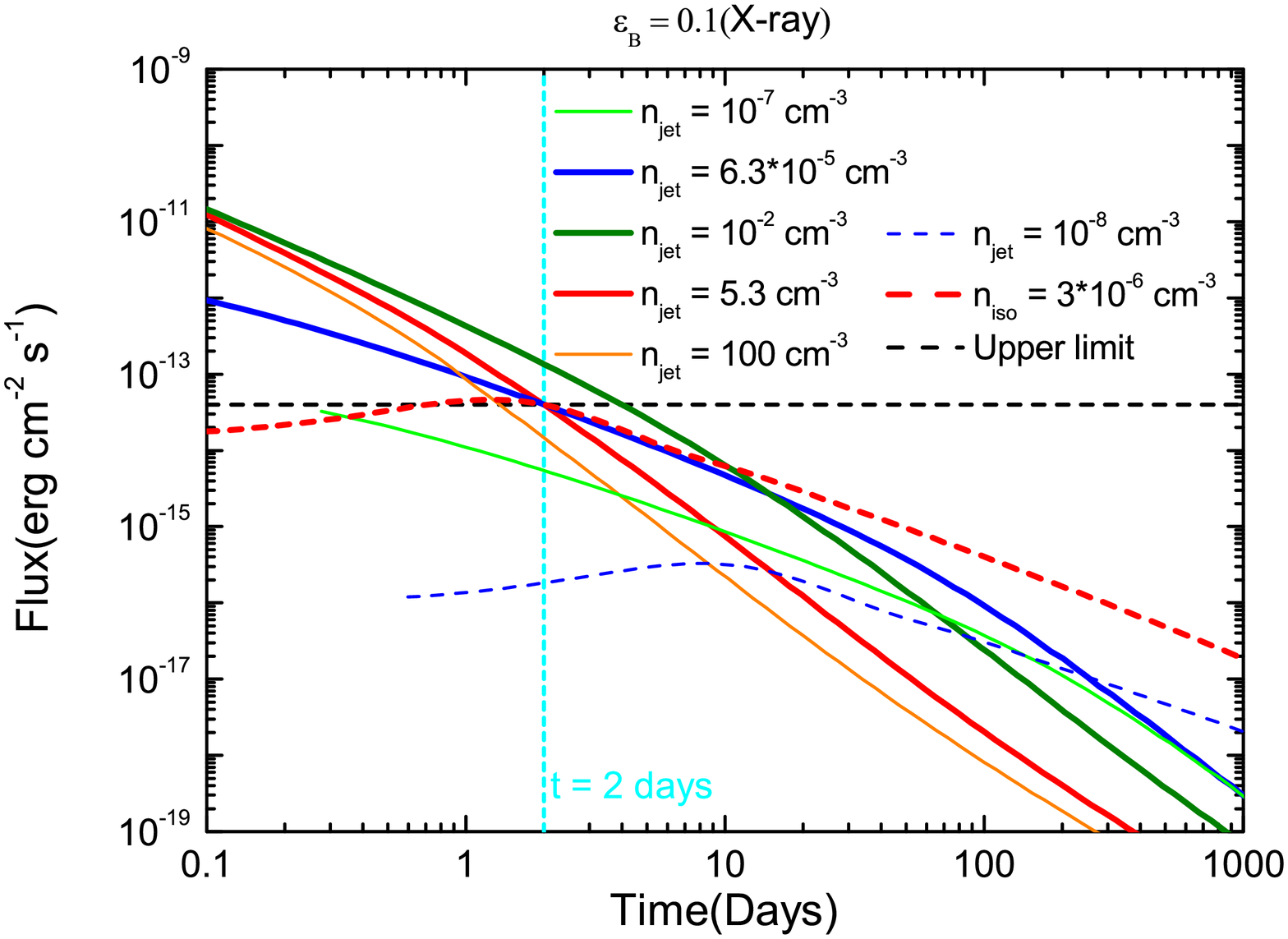} %
\end{center}
\caption{Same as in Fig. \ref{fig:0.01} but $\epsilon_B$ is 0.1.}
\label{fig:0.1}
\end{figure}

The parameter limits derived from the unified dynamic model are listed in Table \ref{tab:1}
and the theoretical light curves are plotted in Fig.~\ref{fig:0.0001}--\ref{fig:0.1}.
We emphasize that the $U$-band emission is fainter than
observations even if $\varepsilon_{\rm B} = 0.1$ and $n = 1$ cm$^{-3}$, indicating that
the upper limit of $U$-band observations cannot be used to constrain the relevant parameters.


\begin{table*}
\begin{center}
\caption{Parameter limits derived from the unified dynamic model}
\label{tab:1}
\begin{tabular}{cccccccccccc}
\hline
\hline
$\varepsilon_{\rm B}$  & $n$(cm$^{-3}$) & $n$(cm$^{-3}$) & $n$(cm$^{-3}$) & $n$(cm$^{-3}$)  \\
                       &   5.5 GHz      &   7.5 GHz      &   X-ray        &   Together    \\
\hline
\hline
{\bf Jet}\\
\hline
     0.0001 & non & non & non &non \\
	 0.001  & $<6.5\times 10^{-2}$ & $<1.6\times 10^{-1}$ &       non            & $<6.5\times 10^{-2}$ \\
	 0.01   & $<6.6\times 10^{-3}$ & $<1.3\times 10^{-2}$ & $<7.4\times 10^{-4}$ & $<7.4\times 10^{-4}$\\
	        &                      &                      &     or $>1.9$        &                     \\
	 0.1    & $<7.4\times 10^{-4}$ & $<1.3\times 10^{-3}$ & $<6.3\times 10^{-5}$ & $<6.3\times 10^{-5}$\\
	        &                      &                      &       or $>5.3$      &                     \\
\hline
\hline
{\bf Isotropic}\\
\hline
     0.0001 & $<6.5\times 10^{-2}$ & $<1.3\times 10^{-1}$ & $<1.0$ & $<6.5\times 10^{-2}$ \\
	 0.001  & $<9.0\times 10^{-3}$ & $<1.0\times 10^{-2}$ & $<1.0\times 10^{-2}$ & $<9.0\times 10^{-3}$ \\
	 0.01   & $<1.0\times 10^{-3}$ & $<1.5\times 10^{-3}$ & $<2.0\times 10^{-4}$ & $<2.0\times 10^{-4}$ \\
	 0.1    & $<8.5\times 10^{-5}$ & $<1.0\times 10^{-4}$ & $<3.0\times 10^{-6}$ & $<3.0\times 10^{-6}$ \\
\hline
\hline
\end{tabular}
\end{center}
\end{table*}

It should be noted that some papers (e.g., \citealt{Dai2016a}, \citealt{Mur2017},
and \citealt{Gao2017}) have constrained the ISM density for the isotropic outflow case.~
For example, \citet{Dai2016a} used the upper limit of the X-ray observations
and inferred that $n\lesssim 2.6\times 10^{-4}$ cm$^{-3}$ if $\varepsilon_{B} = 0.01$,
being roughly consistent with the value inferred here ($2.0\times 10^{-4}$ cm$^{-3}$).

According to Table \ref{tab:1}, the ISM density derived by the
model that takes into account the jet effect and the non-relativistic effect is apparently
different from that derived by the isotropic models adopted by previous studies
(e.g., \citealt{Dai2016a}, \citealt{Mur2017}, and \citealt{Gao2017}). Moreover,
the constraint on the ISM density obtained by multi-band upper limits is tighter
than that derived by any single band. These two facts demonstrate the necessity of employing
the unified model and combining the data at all bands.

We give Table \ref{tab:2} to describe the properties adopted by four models
(this paper, \citealt{Dai2016a}, \citealt{Mur2017}, and \citealt{Gao2017}) and their
results.


\begin{table*}
\begin{center}
\caption{The differences of our model and three previous models}
\label{tab:2}
\begin{tabular}{cccccccccccc}
\hline
\hline
       & this paper & Murase17 & Gao17 & Dai16  \\
\hline
\hline
Isotropic  & yes & yes &  yes  & yes \\
Jet  &yes  &no  &no  &no  \\
Relativistic   & yes & yes & yes & yes\\
Non-relativistic   & yes & no & no & no\\
kilonova &yes & no & no & no \\
X-ray   &yes & no & yes & yes \\
Optical  &yes & no & no & no \\
Radio    &yes & yes & yes & yes \\
\hline
\hline
\end{tabular}
\end{center}
\end{table*}

\subsection{The contours}

Fig.~\ref{fig:con} consists of four contour sub-figures corresponding to the detected
flux at 5.5 GHz ($t = 3$ days), 7.5 GHz ($t = 3$ days), $U$-band
($t = 2$ days), and X-ray ($t = 2$ days). The red lines represent the detection
limits. Every panel has $48 \times 48$ data points each of which represents
the value of $\varepsilon_{\rm B}$ and $n$ at this point.
The curves in these panels represent the detection limits, and
the admitted (two dimensional) parameter space is lower than the lines.
Panel c (corresponding to $U$-band) has no line, indicating that the
theoretical $U$-band flux reproduced by all possible parameters are
smaller than the detected limit and cannot pose any constraint on the
parameters.

\section{Discussion}
\label{sec:dis}

\subsection{The possible origin of Swift~J0644.5-5111/FRB~131104}

It has long been believed that the GRB prompt emission would be
produced in the relativistic jet launched by the core collapse of massive stars
\citep{Uso1992,Woos1993,Thom1995,Mac1999}
or mergers of compact star systems containing at least one neutron star.
The GRBs are usually divided into two categories: long GRBs with
$T_{90}>2$ s are linked to the core collapse of massive stars while short GRBs with
$T_{90}<2$ s are linked to the mergers of binary compact stars.
However, some long GRBs (e.g., GRB~060614, \citealt{Yang2015})
share the properties of short GRBs and can be classified to ``long-short bursts"
which might be produced by NS-NS mergers or BH-NS mergers.
The massive star scenario can be excluded both because the required medium
density is very low (as in section \ref{sec:fit}) and because the association
of Swift~J0644.5-5111/FRB~131104 requires that the FRB was emitted before
the gamma-ray radiation but the scenario gives the reverse time sequence
\citep{Gao2017}. Therefore, the possibility that Swift~J0644.5-5111 is
a ``long-short burst" produced by an NS-NS/BH merger event must be investigated.
\citet{Mur2017} have suggested that Swift J0644.5-5111 might
be associated with a gamma-ray burst and discussed various possible origins of Swift J0644.5-5111
(including the possibility that it might be associated with an NS-NS/BH merger). Here,
we discuss the possible origin of Swift~J0644.5-5111/FRB~131104 with a different approach.

A long-short burst might be powered by an NS-NS merger event or
an BH-NS merger event\citep{Yang2015}. While the former cannot be excluded, we
discuss the latter that has also been proposed to be a possible
scenario producing FRBs with luminosity $\sim 10^{40}-10^{41}$
erg\,s$^{-1}$ \citep{Min2015}. In the BH-NS merger model, the
accretion disk is a cold thin disk \citep{Ros2005} and the
duration of the GRB is approximately equal to the viscosity
timescale $t_{\rm vis}$ of the accretion disk
\begin{equation}
t_{\rm vis}=\frac{1}{\alpha \Omega_{\rm k}}\left( \frac{R_{\rm d}}{H}\right) ^{2}\simeq
274~M_{{\rm BH},\text{1}}^{-\frac{1}{2}}R_{\rm d,8}^{\frac{3}{2}}\alpha
_{-2}^{-1}\left( \frac{h}{0.1}\right)^{-2}~{\rm s},
\end{equation}
where $\alpha \sim 0.01-0.1$ \citep{Llo2016} is the viscosity of the accretion disk,
$\Omega_{k}=\sqrt{GM_{\rm BH}/R_{\rm d}^3}$~is the Keplerian velocity of the accretion
disk, $G$~is the gravitational constant, $M_{\rm BH}=M_{\rm BH,1}\times 10M_\odot$
is the mass of the post-merger BH, $H=h\times R_{\rm d}$ is the scale-height of the
disk, and $R_{\rm d}=R_{\rm d,S}\times R_{\rm S}$ is the outer radius of the disk
with $R_{\rm S}$ being the Schwarzschild radius of the BH. The numerical simulations
performed by \citet{Ros2004} have shown that the radius of the outer disk can be as
large as 600~km.

Letting $h\equiv$ $H/R_{\rm d}$ and adopting $H=0.1R_{\rm d}$, the estimated viscosity
timescale $t_{\rm vis}$ can be approximately equal to the duration of Swift~J0644.5-5111.
Therefore, we suggest that Swift~J0644.5-5111 might be a long-short burst originating from
a NS-BH merger event that could also have produced FRB~131104.

While FRB 131104 might be powered by a NS-BH merger event, we note
that only a minor fraction of FRBs can be supposed to be associated with an
NS-NS/BH merger events since the FRB rate is much higher than the predictions
for compact object coalescence rate. In other words, FRB 131104 would have to
be fundamentally different than other FRBs.

\subsection{The contribution from the UV emission of a kilonova}

The merger of a NS-NS binary or a BH-NS binary would
eject $\sim0.001-0.1M_\odot$ of neutron-rich material and
synthesize some $r$-process heavy elements.
The radioactive decay of these $r$-process elements
could heat the ejecta and yield UV-optical-IR radiation (named  ``kilonova"
e.g., \citealt{Li1998,Met2010,BK13}).

\begin{figure}[tbph]
\begin{center}
\includegraphics[width=0.8\textwidth,angle=0]{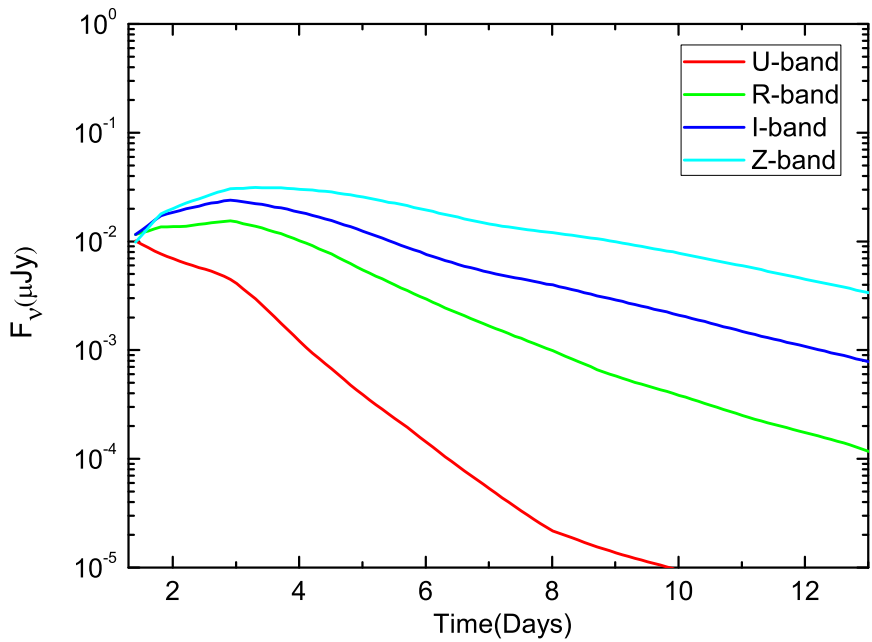} %
\end{center}
\caption{The light curves of a kilonova at a redshift of $0.55$.}
\label{fig:kn}
\end{figure}

In the BH-NS merger scenario, the contribution from the emission of
a kilonova must be taken into account. We use the equations derived
by \citet{Kaw2016} to calculate the multi-band emission from a kilonova.
The parameters adopted are listed below:
the optical opacity $\kappa = 10$ cm$^2$~g$^{-1}$,
the ejecta mass $M_{\rm ej} = 0.02 M_\odot$,
the average velocity of the mass $v_{\rm ej} = 0.18 c$,
the minimum velocity of the mass $v_{\rm ej,min} = 0.02 c$,
the efficiency of thermalization $\epsilon_{\rm th} = 0.5$.
The multi-band ($U$, $R$, $I$, and $Z$) light curves are plotted in
Fig.~\ref{fig:kn}. We find that the $U$-band emission ($<10^{-2}$ $\mu$Jy) is
significantly lower than the upper limit ($\sim$ 30 $\mu$Jy) of the $U$-band
observation and can be neglected.

\section{Conclusions}
\label{sec:con}

The long-duration ($T_{90}\approx$ 377 s) gamma-ray transient Swift~J0644.5-5111
was supposed to be associated with FRB~131104~\citep{DeL2016}. The follow-up
observations for Swift~J0644.5-5111 got the upper limits of radio (5.5 GHz
and 7.5 GHz), $U$-band and X-ray radiation.
Previous research used the standard GRB afterglow model to constrain the
properties of the ISM surrounding the progenitor system of
Swift~J0644.5-5111/FRB~131104. These works have neglected the jet effect
and the jet evolution effect.

In this paper, we have adopted the unified dynamic model
for the multi-band afterglows of GRBs and considered the upper limits of the
observations for the electromagnetic counterparts at 5.5 GHz, 7.5 GHz,
$U$-band, and X-ray to constrain two most important parameters,
$\varepsilon_{\rm B}$ and $n$.
To explain how our models are fundamentally different than those
presented elsewhere, we presented Table \ref{tab:2} that can differentiate our model and
the previous models.

Using this model, we found that
when $\varepsilon_{\rm B}=0.0001$, $n$ cannot be constrained (in the jet case)
or $n \leq 6.5\times 10^{-2}$ cm$^{-3}$ (in the isotropic case);
when $\varepsilon_{\rm B}=0.001$, $n \leq 6.5\times 10^{-2}$ cm$^{-3}$ (jet)
or $n \leq 9.0\times 10^{-2}$ cm$^{-3}$ (isotropic);
when $\varepsilon_{\rm B}=0.01$, $n \leq 7.4\times 10^{-4}$ cm$^{-3}$ (jet)
or $n \leq 2.0\times 10^{-4}$ cm$^{-3}$ (isotropic);
when $\varepsilon_{\rm B}=0.1$, $n \leq 6.3\times 10^{-5}$ cm$^{-3}$ (jet)
or $n < 3.0\times 10^{-6}$ cm$^{-3}$ (isotropic). It can be found that
differences between the jet case and isotropic case are so obvious that the
jet effect must be taken into account. If we have more accurate afterglow data,
the multi-band light curves produced by a jet and an isotropic outflow can be
distinguished since the former emission would produce a jet break in light curves.

\begin{figure}[tbph]
\begin{center}
\includegraphics[width=0.35\textwidth,angle=0]{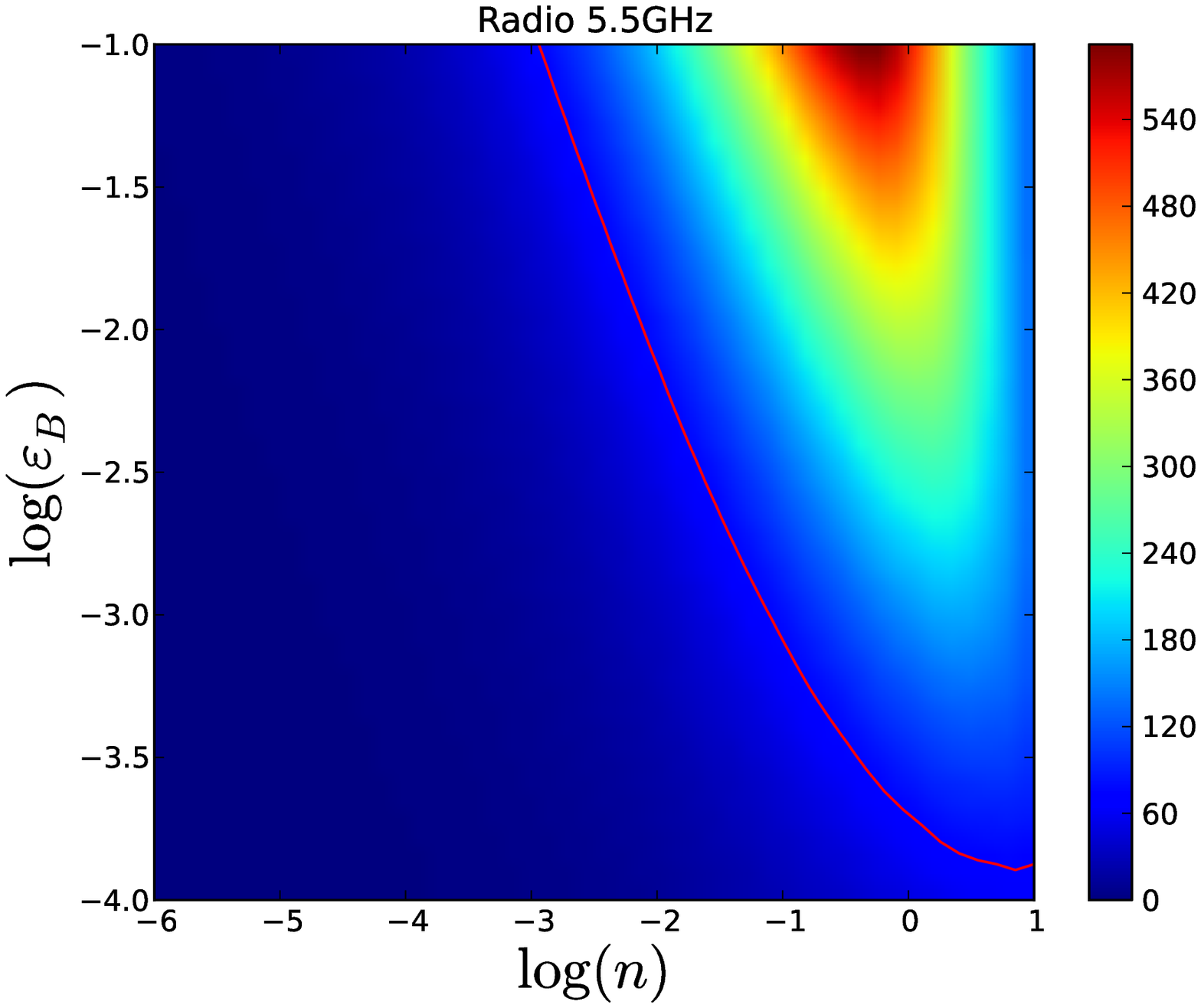} %
\includegraphics[width=0.35\textwidth,angle=0]{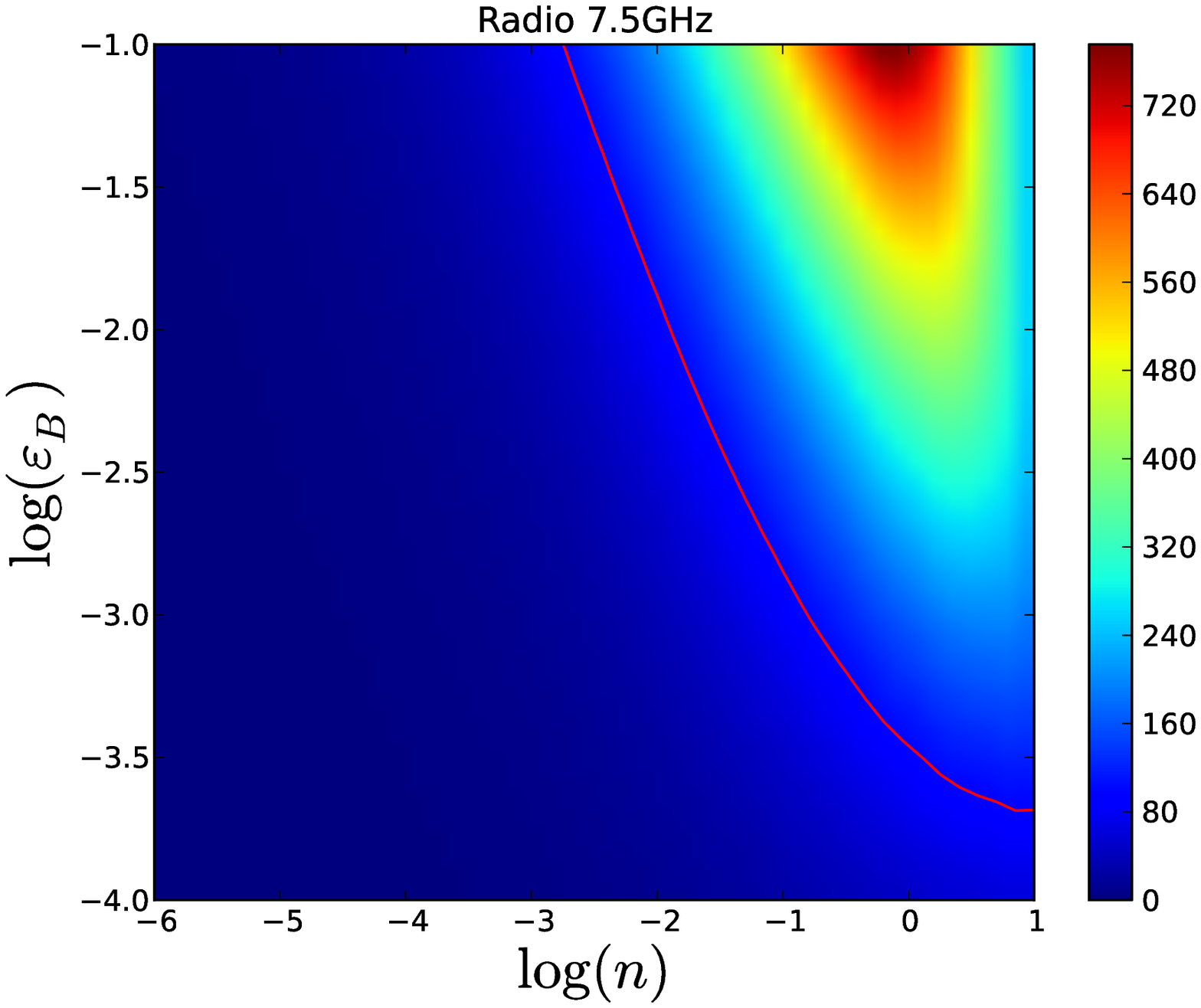}\\[0pt]
\includegraphics[width=0.35\textwidth,angle=0]{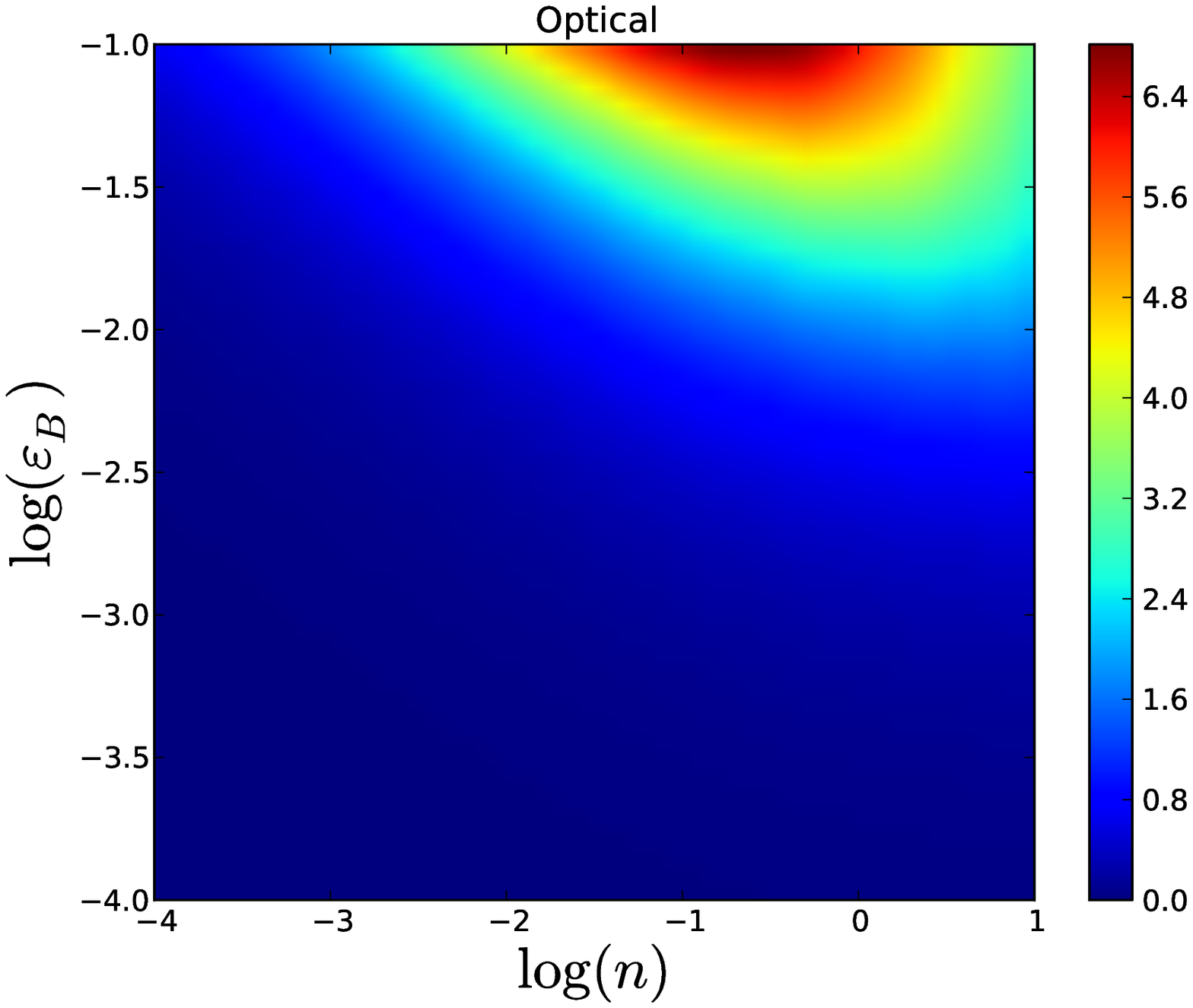} %
\includegraphics[width=0.35\textwidth,angle=0]{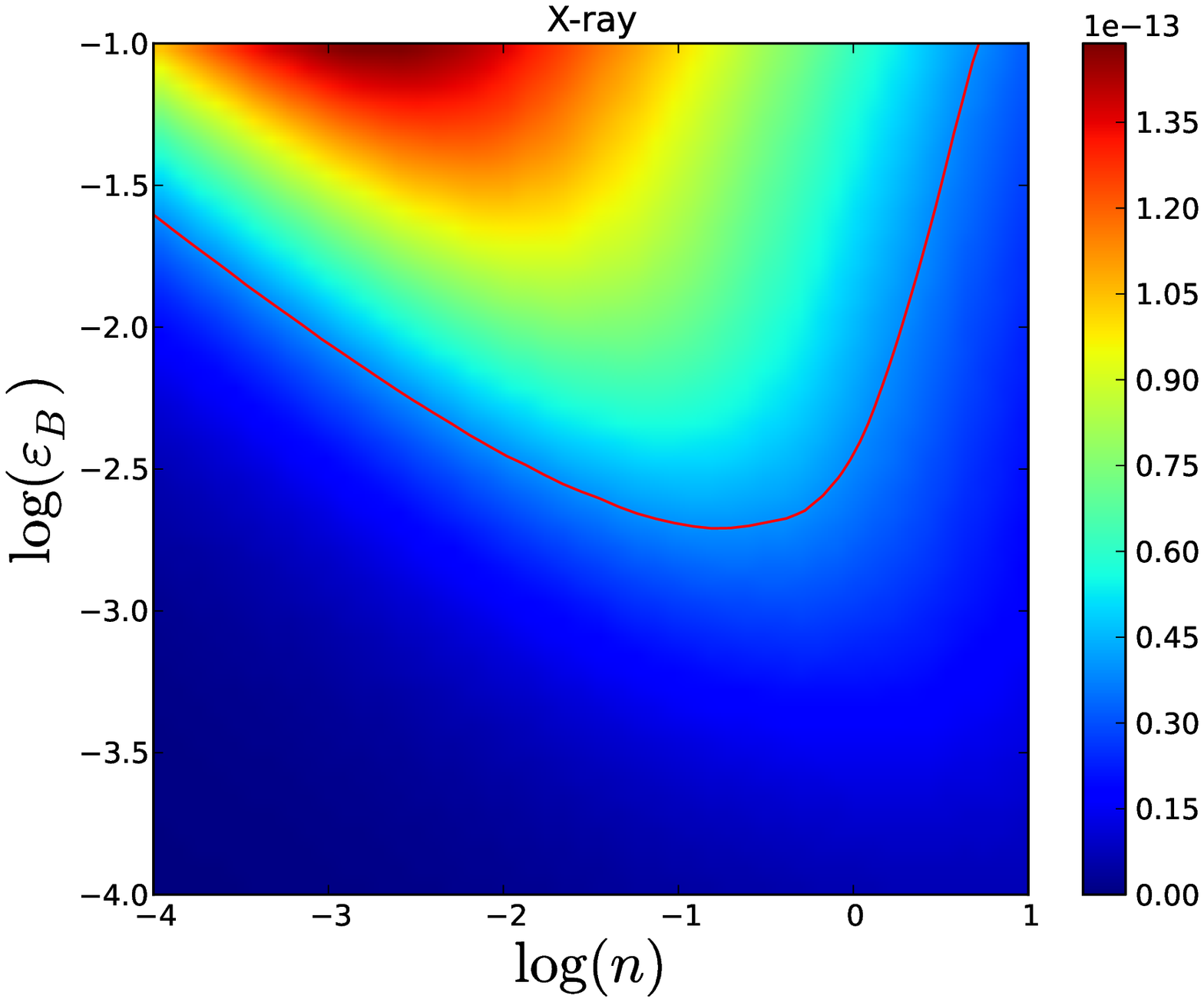}
\end{center}
\caption{Contours for the detected flux at 5.5 GHz, 7.5 GHz,  $U$-band, and
X-ray. The red lines represent the detect limits. Panel c has no line, indicating
that the theoretical $U$-band flux reproduced by all possible parameters are smaller
than the detected limit.}
\label{fig:con}
\end{figure}

Furthermore, we plotted the $n$--$\varepsilon_{\rm B}$ contour sub-figures
(see Fig.~\ref{fig:con}) for the upper limits at 5.5 GHz,
7.5 GHz, $U$-band, as well as X-ray, and found that the inferred upper
limits of the density of the environment of the progenitor are
consistent with both that of massive stars ($n \sim 1-10$ cm$^{-3}$)
and neutron stars ($n \ll $ 1 cm$^{-3}$ and can be as low as
$10^{-8}$ cm$^{-3}$, e.g., \citealt{Zheng2014}).
Since the duration $T_{90}\approx 377$ s and the massive-star
scenario is excluded, the remaining possibilities are
that Swift~J0644.5-5111 is an off-axis short GRB with
an extended emission \citep{Dai2016a} or a long-short burst.
Besides, the results above suggest that the
environmental properties of Swift~J0644.5-5111/FRB 131104 favor a
BH-NS merger scenario, supporting the conclusion of \citet{Mur2017}.

Since the BH-NS merger must eject $0.001-0.1 M_\odot$ of neutron-rich material
and a kilonova powered by the radioactivity of $r$-process elements synthesized
in it must be taken into account, we calculated the multi-band emission from
the kilonova. By comparing the $U$-band luminosity of the kilonova and the upper limit
of the U-band observation, we found that the former is significantly lower than
the latter. Therefore we can conclude that the kilonova emission can be neglected
and the constraints obtained in section \ref{sec:fit} are valid without further
modifications.

We caution that the connection between the Swift~J0644.5-5111/FRB 131104 is tenuous and
all other FRBs do not have gamma-ray counterparts. Based on the large distance inferred
from the DM of FRB 131104 and the fact that Swift~J0644.5-5111 is a highly off-axis
event, the derived isotropic gamma-ray energy of Swift~J0644.5-5111 is very high.
It would be expected that upcoming multi-band and multi-messenger observations
for FRBs could pose more stringent constraints on all relevant parameters,
shedding more light on the nature of FRBs. \footnote{It should be noted that the
distances of FRBs are orders of magnitude larger than the distance limits of detections
of current and near-Future GW detectors so that the rate of
GW events associated with NS-NS/BH mergers in general is much lower than the FRB rate.}

\acknowledgments We thank an anonymous referee for helpful comments and
suggestions that have allowed us to improve our manuscript.
We thank Yong-Feng Huang, Jin-Jun Geng, and Long-Biao Li
for helpful discussion.
This work was supported by the National Basic Research Program
(``973" Program) of China (grant no. 2014CB845800) and the
National Natural Science Foundation of China (grant no. 11573014).
S.Q.W., L.D.L., and H.Y. are supported by the China Scholarship
Program to conduct research at UCB, UNLV and KSU, respectively.


\clearpage

\clearpage

\clearpage

\begin{thebibliography}{Fialkov \& Loeb(2017)}
\bibitem[Abadie et al.(2010)]{Aba2010} Abadie, J., Abbott, B.~P., Abbott, R., et al.\ 2010, Classical and Quantum Gravity, 27, 173001
\bibitem[Burke-Spolaor \& Bannister(2014)]{Bur2014} Burke-Spolaor, S., \& Bannister, K.~W.\ 2014, \apj, 792, 19
\bibitem[Caleb et al.(2017)]{Cal2017} Caleb, M., Flynn, C., Bailes, M., et al.\ 2017, \mnras, 468, 3746
\bibitem[Connor et al.(2016)]{Con2016} Connor, L., Sievers, J., \& Pen, U.~L.\ 2016, \mnras, 458, L19
\bibitem[Dai \& Lu(1998a)]{Dai1998a} Dai, Z.~G., \& Lu, T.\ 1998a, \mnras, 298, 87
\bibitem[Dai \& Lu(1998b)]{Dai1998b} Dai, Z.~G., \& Lu, T.\ 1998b, \aap, 333, L87
\bibitem[Dai et al.(2016a)]{Dai2016a} Dai, Z.~G., Wang, J.~S., \& Wu, X.~F.\ 2016a, arXiv:1611.09517
\bibitem[Dai et al.(2016b)]{Dai2016b} Dai, Z.~G., Wang, J.~S., Wu, X.~F., \& Huang, Y.~F.\ 2016b, \apj, 829, 27
\bibitem[Dai et al.(2017)]{Dai2017} Dai, Z.~G., Wang, J.~S., \& Yu, Y.~W.\ 2017, ApJL, 838, L7
\bibitem[DeLaunay et al.(2016)]{DeL2016} DeLaunay, J.~J., Fox, D.~B., Murase, K., et al.\ 2016, ApJL, 832, L1
\bibitem[Falcke \& Rezzolla(2014)]{Fal2014} Falcke, H., \& Rezzolla, L.\ 2014, \aap, 562, A137
\bibitem[Gao \& Zhang(2017)]{Gao2017} Gao, H., \& Zhang, B.\ 2017, ApJL, 835, L21
\bibitem[Geng \& Huang(2015)]{Geng2015} Geng, J.~J., \& Huang, Y.~F.\ 2015, \apj, 809, 24
\bibitem[Gu et al.(2016)]{Gu2016} Gu, W.~M., Dong, Y.~Z., Liu, T., Ma, R., \& Wang, J.\ 2016, ApJL, 823, L28
\bibitem[Hassall et al.(2013)]{Has2013} Hassall, T.~E., Keane, E.~F., \& Fender, R.~P.\ 2013, \mnras, 436, 371
\bibitem[Huang et al.(1999)]{Huang1999} Huang, Y.~F., Dai, Z.~G., \& Lu, T.\ 1999, \mnras, 309, 513
\bibitem[Huang et al.(2000a)]{Huang2000a} Huang, Y.~F., Dai, Z.~G., \& Lu, T.\ 2000a, \mnras, 316, 943
\bibitem[Huang et al.(2000b)]{Huang2000b} Huang, Y.~F., Gou, L.~J., Dai, Z.~G., \& Lu, T.\ 2000b, \apj, 543, 90
\bibitem[Hurley et al.(2005)]{Hur2005} Hurley, K., Boggs, S.~E., Smith, D.~M., et al.\ 2005, \nat, 434, 1098
\bibitem[Kashiyama et al.(2013)]{Kas2013} Kashiyama, K., Ioka, K., \& M{\'e}sz{\'a}ros, P.\ 2013, ApJL, 776, L39
\bibitem[Katz(2014)]{Kat2014} Katz, J.~I.\ 2014, \prd, 89, 103009
\bibitem[Kawaguchi et al.(2016)]{Kaw2016} Kawaguchi, K., Kyutoku, K., Shibata, M., \& Tanaka, M. 2016, \apj, 825, 52
\bibitem[Keane et al.(2016)]{Kea2016} Keane, E.~F., Johnston, S., Bhandari, S., et al.\ 2016, \nat, 530, 453
\bibitem[Kulkarni et al.(2014)]{Kul2014} Kulkarni, S.~R., Ofek, E.~O., Neill, J.~D., Zheng, Z., \& Juric, M.\ 2014, \apj, 797, 70
\bibitem[Li \& Paczy{\'n}ski(1998)]{Li1998} Li, L.~X., \& Paczy{\'n}ski, B.\ 1998, ApJL, 507, L59
\bibitem[Li et al.(2017)]{Li2017} Li, L.-B., Huang, Y.-F., Zhang, Z.-B., Li, D., \& Li, B.\ 2017, Research in Astronomy and Astrophysics, 17, 6
\bibitem[Lingam \& Loeb(2017)]{Lin2017} Lingam, M., \& Loeb, A.\ 2017, ApJL, 837, L23
\bibitem[Lloyd-Ronning et al.(2016)]{Llo2016} Lloyd-Ronning, N.~M., Dolence, J.~C., \& Fryer, C.~L.\ 2016, \mnras, 461, 1045
\bibitem[Lorimer et al.(2007)]{Lor2007} Lorimer, D.~R., Bailes, M., McLaughlin, M.~A., Narkevic, D.~J., \& Crawford, F.\ 2007, Science, 318, 777
\bibitem[MacFadyen \& Woosley(1999)]{Mac1999} {MacFadyen}, A. I., \& {Woosley}, S. E. 1999, \apj, 524, 262
\bibitem[M{\'e}sz{\'a}ros \& Rees(1997)]{Mes1997} M{\'e}sz{\'a}ros, P., \& Rees, M.~J.\ 1997, \apj, 476, 232
\bibitem[Metzger et al.(2010)]{Met2010}Metzger, B. D., Martinez-Pinedo, G., Darbha, S., et al. \mnras, 2010, 406, 2650
\bibitem[Barnes \& Kasen(2013)]{BK13} Barnes, J., \& Kasen, D. \apj, 2013, 775, 18
\bibitem[Mingarelli et al.(2015)]{Min2015} Mingarelli, C.~M.~F., Levin, J., \& Lazio, T.~J.~W.\ 2015, ApJL, 814, L20
\bibitem[Mottez \& Zarka(2014)]{Mot2014} Mottez, F., \& Zarka, P.\ 2014, \aap, 569, A86
\bibitem[Murase et al.(2017)]{Mur2017} Murase, K., M{\'e}sz{\'a}ros, P., \& Fox, D.~B.\ 2017, ApJL, 836, L6
\bibitem[Palaniswamy et al.(2018)]{Pal2018} Palaniswamy, D., Li, Y., \& Zhang, B.\ 2018, \apjl, 854, L12
\bibitem[Piro \& Burke-Spolaor(2017)]{Pir2017} Piro, A.~L., \& Burke-Spolaor, S.\ 2017, ApJL, 841, L30
\bibitem[Ravi et al.(2016)]{Rav2016} Ravi, V., Shannon, R.~M., Bailes, M., et al.\ 2016, Science, 354, 1249
\bibitem[Rosswog(2005)]{Ros2005} Rosswog, S.\ 2005, \apj, 634, 1202
\bibitem[Rosswog et al.(2004)]{Ros2004} Rosswog, S., Speith, R., \& Wynn, G.~A.\ 2004, \mnras, 351, 1121
\bibitem[Santana et al.(2014)]{San2014} Santana, R., Barniol Duran, R., \& Kumar, P.\ 2014, \apj, 785, 29
\bibitem[Sari et al.(1998)]{Sar1998} Sari, R., Piran, T., \& Narayan, R.\ 1998, ApJL, 497, L17
\bibitem[Shannon \& Ravi(2017)]{Sha2017} Shannon, R.~M., \& Ravi, V.\ 2017, ApJL, 837, L22
\bibitem[Thompson \& Duncan(1995)]{Thom1995}Thompson, C., \& Duncan, R. C. 1995, \mnras, 275, 255
\bibitem[Thornton et al.(2013)]{Tho2013} Thornton, D., Stappers, B., Bailes, M., et al.\ 2013, Science, 341, 53
\bibitem[Totani(2013)]{Tot2013} Totani, T.\ 2013, \pasj, 65, L12
\bibitem[Usov(1992)]{Uso1992}{Usov}, V.~V. 1992, \nat, 357, 472
\bibitem[Wang et al.(2016)]{Wang2016} Wang, J.~S., Yang, Y.~P., Wu, X.~F., Dai, Z.~G., \& Wang, F.~Y.\ 2016, ApJL, 822, L7
\bibitem[Williams \& Berger(2016)]{Wil2016} Williams, P.~K.~G., \& Berger, E.\ 2016, ApJL, 821, L22
\bibitem[Woosley(1993)]{Woos1993} {Woosley}, S. E. 1993, \apj, 405, 273
\bibitem[Yang et al.(2015)]{Yang2015} Yang, B., Jin, Z.~P., Li, X., et al.\ 2015, Nature Communications, 6, 7323
\bibitem[Yang \& Zhang(2017)]{Yang2017} Yang, Y.-P., \& Zhang, B.\ 2017, arXiv:1712.02702
\bibitem[Zadorozhna(2015)]{Zad2015} Zadorozhna, L.~V.\ 2015, Advances in Astronomy and Space Physics, 5, 43
\bibitem[Zhang(2016)]{Zhang2016} Zhang, B.\ 2016, ApJL, 827, L31
\bibitem[Zheng et al.(2014)]{Zheng2014} Zheng, Z., Ofek, E.~O., Kulkarni, S.~R., Neill, J.~D., \& Juric, M.\ 2014, \apj, 797, 71


\end{thebibliography}
\end{document}